\documentclass[12pt,hidelinks,a4paper]{article}

\usepackage[english]{babel}
\usepackage[margin=0.8in]{geometry}
\usepackage{lipsum}
\usepackage{setspace}
\usepackage{lineno}
\usepackage{amsmath,amssymb,amsfonts,mathtools}%
\usepackage{natbib}
\usepackage{orcidlink}
\usepackage{hyperref}
\usepackage{palatino}

\DeclareMathOperator*{\argminB}{argmin} 

\usepackage{authblk}
\title{Modelling higher education dropouts using sparse and interpretable post-clustering logistic regression}
\author[1]{Andrea Nigri\footnote{Corresponding author: \texttt{andrea.nigri@unifg.it}.}}
\author[2]{Massimo Bilancia}
\author[1]{Barbara Cafarelli}
\author[3]{Samuele Magro}
\affil[1]{Department of Economics, Management and Territory Department (DEMeT), University of Foggia, Foggia, Italy}
\affil[2]{Department of Precision and Regenerative Medicine and Jonian Area (DiMePRe-J), 
	University of Bari Aldo Moro, Bari, Italy}
\affil[3]{Sirio Astronomical Observatory, Grotte di Castellana srl, Castellana Grotte, Italy}
\date{}                     
\date{\today}

\begin{document}

\maketitle
\begin{abstract}
	
Higher education dropout constitutes a critical challenge for tertiary education systems worldwide. While machine learning techniques can achieve high predictive accuracy on selected datasets, their adoption by policymakers remains limited and unsatisfactory, particularly when the objective is the unsupervised identification and characterization of student subgroups at elevated risk of dropout. The model introduced in this paper is a specialized form of logistic regression, specifically adapted to the context of university dropout analysis. Logistic regression continues to serve as a foundational tool among reliable statistical models, primarily due to the ease with which its parameters can be interpreted in terms of odds ratios. Our approach significantly extends this framework by incorporating heterogeneity within the student population. This is achieved through the application of a preliminary clustering algorithm that identifies latent subgroups, each characterized by distinct dropout propensities, which are then modeled via cluster-specific effects.
We provide a detailed interpretation of the model parameters within this extended framework and enhance interpretability by imposing sparsity through a tailored variant of the LASSO algorithm. To demonstrate the practical applicability of the proposed methodology, we present an extensive case study based on the Italian university system, in which all the developed tools are systematically applied.
	
\end{abstract}

\singlespacing
 
\section{Introduction}

Higher education dropout, defined as the cessation of studies before the successful completion of a degree program, represents a significant challenge within tertiary education systems globally. While the precise definition can vary across contexts, often encompassing students who do not enroll for consecutive academic periods or fail to graduate without formal sanctions, understanding its multifaceted nature is crucial. This failure to complete higher education not only represents a loss of potential human capital but also signifies a considerable drain on the resources invested in transformartive educational endeavors \citep{bilquise_ai-based_2022}. The decision to leave higher education prematurely carries significant consequences that extend across individual, institutional, and societal levels \cite{cherif_understanding_2020}. At the individual level, dropping out can have a detrimental impact on self-esteem and overall mental well-being . Students may experience feelings of inadequacy, self-doubt, and a diminished sense of life satisfaction. These negative effects can extend beyond the immediate academic context, contributing to reduced job satisfaction and even poorer health outcomes in the long term  \citep{litalien_motivation_2015}. Higher education institutions also bear the consequences of student dropout. Economically, high attrition rates can lead to substantial financial losses due to decreased enrollment and the associated reduction in tuition revenue. Academically, low completion rates can negatively impact an institution's reputation and its ability to attract prospective students.
In response, the global academic community is actively developing retention strategies to support students in completing their degrees. Student retention rates and graduation timelines serve as key indicators of both individual success and institutional effectiveness, while also contributing to broader economic and social progress \citep{bilquise_ai-based_2022}. The decision to drop out is not random but rather the outcome of a complex process influenced by multiple factors. These factors can generally be classified into three main categories: national education system-related aspects, institutional policies and environmental conditions, and individual student characteristics \citep{behr_dropping_2020}.  

The persistent issue of student attrition has prompted extensive research into the various factors influencing dropout rates. While traditional statistical methods remain valuable, they often fail to capture the intricate interactions among the variables that shape a student's decision to discontinue their studies \citep{da_costa_dropout_2018,vaarma_predicting_2024}. In contrast, machine learning offers a promising approach to addressing this challenge by employing algorithms capable of detecting complex patterns and improving the accuracy of dropout predictions \citep{rabelo_model_2025}. By analyzing large datasets encompassing academic performance, socioeconomic background, engagement metrics, and institutional factors, machine learning techniques can identify subtle indicators of potential dropout. In this context, the development of supervised predictive models facilitates the early identification of at-risk students, enabling institutions to intervene proactively and provide targeted support \citep{sansone_beyond_2019}. This approach enhances the understanding of the factors contributing to student attrition, ultimately informing the design of more effective interventions and support systems. A related yet distinct challenge lies in the unsupervised identification and characterization of student subgroups at heightened risk of dropout \citep{mohamed_nafuri_clustering_2022}. 

Although the use of machine learning techniques allows for the achievement of high accuracy on selected datasets when the model is applied in a predictive context, the overall adoption of these tools by policymakers remains limited and unsatisfactory. A primary factor behind this challenge is the lack of interpretability--an abstract property of a model that provides insight into which subsets of inputs are most influential in determining the outcomes, or how alterations to an input would affect the result \citep{marcinkevics_interpretable_2023,ortigossa_explainable_2024}. It is also highly desirable that the proposed model be inherently interpretable, meaning that its internal processes are readily comprehensible to policymakers. This additional characteristic fosters greater confidence in the decisions recommended by the model, ensuring that the results are credible, trustworthy, and can be generalized or applied across different contexts or settings \citep{molnar2024}.

The model proposed in this paper can essentially be regarded as a logistic regression framework tailored to the analysis of university dropout. Logistic regression remains a cornerstone among reliable statistical models, owing to the straightforward interpretation of its parameters in terms of odds ratios. However, our approach substantially extends this framework by accounting for heterogeneity within the student population, identified through a preliminary clustering algorithm applied to the data.
For each input variable, a cluster-specific effect--reflecting the propensity to drop out within a given sub-population--is captured through the estimation of an interaction term between the input variable and the cluster label. This formulation renders the conventional interpretation of odds ratios infeasible, as meaningful interpretation becomes possible only conditional on the cluster label.
To enhance interpretability, we introduce sparsity into the model by employing a tailored version of the group-LASSO regularization algorithm--a variant of the standard LASSO specifically designed to accommodate first-order interactions under a strong hierarchy constraint-- whereby an interaction term may be included only if the corresponding main effect is present \citep{lim_learning_2015}. The parameterization adopted by the group-LASSO algorithm departs from the standard form, relying instead on a sum-to-zero constraint. For this reason, we also provide a detailed explanation of how the newly defined cluster-based odds ratios can be computed under this alternative parameterization, thereby substantially extending the results presented in \cite{bilancia_interpretable_2024}.

The paper is organized as follows. Section \ref{sec:methods} introduces the necessary notation and presents the definition of the logistic regression model with interactions, including the interpretation of odds ratios computed conditionally on the cluster labels. In particular, Subsection~\ref{sec:estimation} discusses a specialized version of the group-LASSO algorithm designed to obtain sparse parameter estimates while respecting the strong hierarchy principle. We also provide a detailed explanation of how to compute odds ratios under the original parameterization of the algorithm, without the need to re-estimate a new logistic regression model restricted to the selected variables.
Section \ref{sec:bootstrap} addresses bootstrap-based inference for the sparse model, since classical inference procedures are not valid after LASSO variable selection, and the resulting $p$-values are therefore unreliable under standard frequentist theory.
Section \ref{sec:example} presents a comprehensive case study based on the Italian university system, in which we apply all the methodological tools developed in the theoretical part of the paper and demonstrate their practical relevance.
Finally, Section \ref{sec:discussion} discusses the strengths and limitations of the proposed approach and outlines potential directions for future research.

\section{Methods}\label{sec:methods}
\subsection{Notation}\label{sec:notation}

Let $(y, \pmb x^\top)$ denote a random training pair observed from a data-generating process (DGP), where $y \in \{0,1\}$ and $\pmb x \in \mathbb{X} \subset \mathbb{R}^p$. The variable $\pmb x$ is referred to as the input variable, representing the features associated with each student, while $y$ denotes the output or response variable. Consider the training set $\mathcal{D} = \{ (y_i, \pmb{x}_i^\top) ; i = 1,\ldots,n \}$, which consists of $n$ historical data points corresponding to students who have either dropped out or completed their university studies. Each pair $(y_i, \pmb{x}_i^\top)$ is assumed to be independently drawn from the random pair $(y, \pmb x^\top)$ according to the underlying DGP.

As is customary, the feature vectors are arranged row-wise in the matrix $\pmb X = (\pmb{x}_1, \ldots, \pmb{x}_n)^\top \in \mathbb{R}^{n \times p}$, providing a representation centered on individual data points. However, when a representation centered on individual variables is required, we write $\pmb{X} = (\pmb{X}_1, \ldots, \pmb{X}_p)$, where $\pmb{X}_j \in \mathbb{R}^{n \times 1}$ denotes the column vector containing the values assumed by the $j$-th variable for $j=1, \ldots, p$. If $\pmb{X}_j$ consists of values from a discrete variable, depending on the parameterization employed, it is often necessary to expand $\pmb{X}_j$ by replacing it with appropriate auxiliary variables to ensure the identifiability of the model. This crucial aspect will be revisited in detail in Section \ref{sec:estimation}. In general, uppercase notation serves to distinguish the column vector $\pmb {X}_j$, which contains the values of a single variable, from the row vector $\pmb{x}_i^\top$, which represents a feature vector spanning $p$ variables. Following the same convention but omitting the bold font, $X_j$ serves as an identifier for the $j$-th variable within the feature vector, disregarding the individual observations, and $X = (X_1, \ldots, X_p)$. Similarly, the binary outcome observations are collected in the vector $\pmb{Y} = (y_1, \ldots, y_n)^\top \in \mathbb{R}^{n \times 1}$, whereas $Y$ is an identifier of the same variabile.

Part of the input variables in the specialized supervised model proposed in Section \ref{sec:methods} are derived from the output of a generic clustering algorithm. At this stage, the same input and outcome variables are used, although they reside in the same space for the purpose of unsupervised searching for the optimal partition. Therefore, each data point is redefined as $(y_i, \pmb{x}_i^\top) \equiv \pmb{d}_i^\top = (d_{i1}, \ldots, d_{i(p+1)})$. The data are organized in the matrix $\pmb{D}_n = (\pmb{d}_1, \ldots, \pmb{d}_n)^\top \in \mathbb{R}^{n \times (p+1)}$. Furthermore, we distinguish between continuous and categorical input variables, arranging them so that for $j = 2, \ldots, q$, with $q < p + 1$, the variables are numerical, while for $j = q + 1, \ldots, p + 1$, only categorical variables are present. As before, $d_j$ serves as an identifier denoting the $j$-th variable in this ordered sequence.

From a general perspective, a clustering algorithm defines a mapping $C$ from the space where the data points (rows) in $\pmb{D}_n$ reside onto the set of labels $\{1, \ldots, k\}$, with $k$ fixed. The partition generated by the clustering algorithm corresponds to the inverse image induced by $C$ and is denoted as $C(\pmb{D}_n)$. Any cluster in $C(\pmb{D}_n)$ is a subset of the rows of $\pmb{D}_n$ that does not share any points with the remaining clusters. An example of a procedure for searching for an optimal partition is presented in the case study in Section \ref{sec:example}.

\subsection{The basic model}\label{sec:themodel}

The model employed for interpreting the partition obtained through the clustering algorithm is defined as follows:

\begin{equation}\label{eq:logisticstrata}
	\mathrm{logit}\left( \Pr(Y = 1 \vert X, C(\pmb D_n)\right) = \beta_0 + \sum_{j=1}^p X_j \beta_j + \sum_{s=2}^k
	\mathcal D_s \gamma_{s} + \sum_{j=1}^p\sum_{s=2}^k \Xi_{js}\theta_{js}.
\end{equation}

In this variable-based perspective, we consider a collection of main effects associated with both the input variables and the labels generated by the clustering algorithm. Without loss of generality, in the model \eqref{eq:logisticstrata}, the feature variables $X_j$ are either continuous or at most binary, i.e., $X_j \in \{0,1\}$, except when it is necessary to specify modifications required for cases where $X_j$ is discrete with three or more levels. In such instances, $X_j$ must be replaced by a representation involving at least two or more auxiliary variables to ensure model identifiability. This approach leads to a more parsimonious notation.  
The main effects associated with clustering labels are represented using standard dummy encoding, with $\mathcal C_1$ taken as the reference level. Thus, $(0, \ldots, 0)^{\top} \in \mathbb{R}^{k-1}$ denotes $\mathcal C_1$, while the remaining labels are encoded by the elements $(\mathcal D_2, \ldots, \mathcal D_k)^{\top}$, which belong to the canonical basis of $\mathbb{R}^{k-1}$. Each main effect $X_j$ is accompanied by an interaction term involving the variables representing the partition labels. In what follows, each interaction variable is defined as the product of the corresponding fixed effects $\Xi_{js} \equiv X_j \mathcal{D}_s$, and can be always formally treated as the product of these two terms when computing derivatives or first differences in this variable-based perspective.

If $\phi(u)$ represents the linear predictor on the right-hand side of \eqref{eq:logisticstrata}, then, if $X_j$ is a continuous variable and given that $\mathcal{D}_s \in \{0,1\}$, we have:
$$
\frac{\Delta \frac{\partial \phi(u)}{\partial X_j}}{\Delta \mathcal D_s} = (\beta_j + 1 \cdot \theta_{js}) - (\beta_j + 0 \cdot \theta_{js}) = \theta_{js},
$$  
which demonstrates that in linear models, it is well-founded to interpret $\theta_{js}$ as a unit variation in the response variable when both the main effects vary simultaneously.

Unfortunately, for an intrinsically nonlinear model such as logistic regression, this standard property does not hold. To understand why, we rewrite the model by isolating the terms $X_j$ and $\mathcal{D}_s$ as follows:
\begin{equation*}
\mathrm{logit}\left( \Pr(Y = 1 \vert X, \mathcal C(\pmb D_n) \right) = \beta_0 + X_j \beta_j + 
\mathcal D_s \gamma_{s} + (X_j \mathcal D_s)\theta_{js} + \mathrm{rest},
\end{equation*} 
and in case that $X_j \in \{0,1\}$, we obtain:  
\begin{align}\label{eq:interaction}
	& \frac{\Delta}{\Delta \mathcal D_s} \frac{\Delta}{\Delta X_j} \Pr(Y=1 \vert X_j, \mathcal D_s, \mathrm{rest})=  \nonumber \\
	= & \frac{\Delta}{\Delta \mathcal D_s}  \frac{\Delta}{\Delta X_j}\left[\frac{1}{1+e^{-(\beta_0 + X_j \beta_j + 
			\mathcal D_s \gamma_{s} + (X_j \mathcal D_s)\theta_{js} + \textrm{rest})}}\right] = \nonumber \\
	= &  \frac{\Delta}{\Delta \mathcal D_s} \left[\frac{1}{1 + e^{-(\beta_0 + \beta_j+ \mathcal D_s \gamma_s+ \mathcal D_s\theta_{js} +  \mathrm{rest})}} - 
	\frac{1}{1 + e^{-(\beta_0 + \mathcal D_s \gamma_s+ \mathrm{rest})}}\right] = \nonumber \\
	= & \left[\frac{1}{1 + e^{-(\beta_0 + \beta_j+ \gamma_s + \theta_{js} +  \mathrm{rest})}} -
	\frac{1}{1 + e^{-(\beta_0 + \beta_j+  \mathrm{rest})}}\right. - \nonumber \\
	- & \left.\frac{1}{1 + e^{-(\beta_0 + \gamma_s +  \mathrm{rest})}} + \frac{1}{1 + e^{-(\beta_0 + \mathrm{rest})}}\right],
\end{align}  
with similar calculations holding when $X_j$ is a continuous variable \citep{ai_interaction_2003, norton_computing_2004}. If, for instance, $X_j$ is a categorical variable with three levels, it suffices to replace $X_j$ with the dummy variables $\mathcal{D}_2(X_j), \mathcal{D}_3(X_j)$, conventionally taking the first level as the reference, and to apply the computations used for \eqref{eq:interaction} separately to each dummy variable. The key implication of expression \eqref{eq:interaction} is that the interaction coefficient $\theta_{js}$ cannot be interpreted as the logarithm of an adjusted odds ratio. This arises because a simultaneous local variation in the main effects induces a global change in the response probability that also depends on all other variables included in the model. For the same reason, analyzing the expression of $\frac{\Delta}{\Delta X_j} \Pr(Y=1 \vert X_j, \mathcal D_s, \mathrm{rest})$ reveals that even the $\beta_j$ coefficients associated with the main effects of individual features do not retain their usual interpretation when interaction terms are present in the model.

\subsubsection{Parameter interpretation}\label{subsec:parameter_interpretation}

Borrowing from \cite{ai_interaction_2003}, we conclude, as demonstrated below, that a correct interpretation of the model parameters can be obtained by conditioning on the cluster labels. In the case where $X_j \in \{0,1\}$ is a binary variable:  
\begin{align}\label{eq:logodds}
	&\log \mathsf{OR}(X_j = 1 \;\mathrm{ vs. }\; X_j = 0\vert \mathcal D_s = 1, \mathcal D_s' = 0, s'\neq s \in \{2,\ldots,k\}, \mathrm{rest})  \nonumber \\
	= \;& \mathrm{logit}(X_j = 1 \vert \mathcal D_s = 1, \mathcal D_s' = 0, s'\neq s \in \{2,\ldots,k\}, \mathrm{rest}) \nonumber \\
	- \;& \mathrm{logit}(X_j= 0 \vert \mathcal D_s = 1, \mathcal D_s' = 0, s'\neq s \in \{2,\ldots,k\}, \mathrm{rest})  \nonumber \\
	= \;& \beta_0 + (X_j = 1) \beta_j + (\mathcal D_s = 1) \gamma_{s}  + (X_j = 1 * \mathcal D_s = 1) \theta_{js} + \mathrm{rest} \nonumber \\
	- \;& \left[ \beta_0 + (X_j = 0) \beta_j  + (\mathcal D_s = 1) \gamma_s + (X_j = 0 * \mathcal D_s = 1) \theta_{js} +  \mathrm{rest} \right]  \nonumber \\
	= \; & \beta_0 + \beta_j + \gamma_s + \theta_{js} + \mathrm{rest}- \beta_0 - \gamma_s -  \mathrm{rest} = \beta_j + \theta_{js},
\end{align}  
where, with a slight abuse introduced to simplify the notation, for example:
\begin{align*}
&  \mathrm{logit}(X_j = 1 \vert \mathcal D_s = 1, \mathcal D_s' = 0, s'\neq s \in \{2,\ldots,k\}, \mathrm{rest})  \\
\equiv \;& \mathrm{logit}\left(\Pr(Y = 1\vert X_j = 1, 
\mathcal D_s' = 0, s'\neq s \in \{2,\ldots,k\}, X_j'\textrm{ const.}, j'\neq j \in \{1,\ldots,p\})\right)
\end{align*}
emphasizing that the cluster indicator dummy is constant, while only the feature variable $X_j$ varies in its domain. As an immediate consequence of \eqref{eq:logodds}:  
\begin{equation}\label{eq:outsideOR}
	\mathsf{OR}(X_j = 1 \;\mathrm{ vs. }\; X_j = 0\vert \mathcal D_s = 1, \mathcal D_s' = 0, s'\neq s \in \{2,\ldots,k\}, \mathrm{rest}) = e^{\beta_j + \theta_{js}}.
\end{equation}  
with similar calculations readily showing that:  
\begin{equation}\label{eq:insideOR}
	\mathsf{OR}(X_j = 1 \;\mathrm{ vs. }\; X_j = 0\vert \mathcal D_s = 0, \mathcal D_s' = 0, s'\neq s \in \{2,\ldots,k\}, \mathrm{rest}) = e^{\beta_j}.
\end{equation}  

We can therefore define two distinct odds ratios: a baseline adjusted odds ratio \eqref{eq:insideOR} for instances assigned to the reference cluster and an adjusted odds ratio \eqref{eq:outsideOR} for instances assigned to the cluster indicated by the dummy variable $\mathcal{D}_s$. The multiplicative constant $e^{\theta_{js}}$ can be interpreted as a ratio of (conditional and adjusted) odds ratios:  
\begin{equation}\label{eq:effectratio}
	\mathsf{ROR}(X_j)=\frac{\mathsf{ OR }(X_j = 1 \;\mathrm{ vs. }\; X_j = 0\vert \mathcal D_s = 1, \mathcal D_s' = 0, s'\neq s \in \{2,\ldots,k\}, \mathrm{rest})}{\mathsf{ OR }(X_j = 0 \;\mathrm{ vs. }\; X_j = 0\vert \mathcal D_s = 0, \mathcal D_s' = 0, s'\neq s \in \{2,\ldots,k\}, \mathrm{rest})} = e^{\theta_{js}}.
\end{equation}  

As we will demonstrate in Section \ref{sec:example}, the quantity defined in \eqref{eq:effectratio} provides a useful measure for interpreting the effect size of each feature input variable on the outcome. Naturally, the considerations outlined above remain entirely valid when $X_j$ is a continuous variable. Conversely, if $X_j$ is a categorical variable with $L \geq 3$ levels, it suffices to define $L-1$ dummy variables and compute a ratio of odds ratios for each of these $L-1$ variables.  

\subsection{Sparse parameter estimation}\label{sec:estimation}

Although the proposed model offers clear advantages by allowing the definition of cluster-specific odds ratios, its interpretability is often compromised by the presence of input features that exert a comparable influence on the outcome probability. From this perspective, the use of LASSO regularization with an $L_1$ penalty represents a principled approach that can significantly enhance the model’s practical applicability \citep{tibshirani_regression_1996,zou_adaptive_2006,hastie_statistical_2015,gauraha_introduction_2018}.

To this end, we rewrite the log-likelihood of the model \eqref{eq:logisticstrata} using a different parameterization necessitated by the estimation algorithm used:
\begin{eqnarray}\label{eq:logisticstratalik}
	\ell(\pmb \beta, \pmb \gamma, \pmb \Xi \vert \pmb Y, \pmb X, \pmb C) & = & \left[  \pmb Y^T \left( \beta_0 \pmb 1  + \sum_{j=1}^p \pmb X_j \pmb \beta_j + \pmb C \pmb \gamma + \sum_{j=1}^p \pmb \Xi_j \pmb\theta_j\right) -\right. \nonumber\\
	& - & \left.\pmb 1^\top \log \left( \pmb 1 + \exp\left(\beta_0 \pmb 1  + \sum_{j=1}^p \pmb X_j \pmb \beta_j + \pmb C \pmb \gamma + \sum_{j=1}^p \pmb \Xi_j \pmb\theta_j \right)\right)\right],
\end{eqnarray}
where $\pmb 1$ denotes a vector containing only ones, and the log and exp functions are taken component-wise. In this expression, if variable $X_j$ is a scalar then $\pmb X_j \in \mathbb R^{n \times 1}$ and $\pmb \beta_j \equiv \beta_j$ is a scalar. If $X_j$ is a categorical variable with $L$ distinct levels, then $\pmb X_j$ is an $n \times L$ matrix and $\pmb\beta_j \in \mathbb R^{L}$ (taken as an $L\times 1$ column vector). Each level of $X_j$ is represented via an element of the canonical basis of $\mathbb R^L$. This one-hot encoding introduces obvious non identifiabilities into the model, and will be treated imposing sum-to-zero constraints. Similarly, the clusters indicators are one-hot encoded and collected into the matrix $\pmb C \in \mathbb R^{n\times k}$, with $\pmb \gamma \in \mathbb R^k$. Finally, the interaction matrices are obtained as the columnwise product of the corresponding fixed-effects matrices, that is, $\pmb\Xi_j = \pmb X_j \star \pmb C$, where every column of $\pmb X_j$ is multiplied with every column of $\pmb C$. In this way, if $\pmb X_j \in \mathbb R^{n\times L}$ then $\pmb \Xi_j \in \mathbb R^{n \times (L\cdot k)}$ and $\pmb \theta \in \mathbb R^{(L\cdot k)}$. With this notation, the penalized log-likelihood \eqref{eq:logisticstratalik} with the LASSO penalty takes the form:
\begin{equation}
	\argminB_{\pmb \psi} \tilde\ell(\pmb \psi \vert \pmb Y, \pmb X, \pmb C) + \lambda \vert\vert \pmb\psi \vert\vert_1
\end{equation}
for some $\lambda \geq 0$, where $\tilde\ell(\pmb\psi \vert \pmb Y, \pmb X, \pmb C) = - \ell(\pmb\psi \vert \pmb Y, \pmb X, \pmb C)$, $\pmb\psi$ collectively represents all model parameters, and $\lVert \cdot \rVert_1$ denotes the $L_1$-norm.

The standard LASSO approach encounters significant limitations when applied to logistic models with interaction terms. A fundamental principle in model construction dictates that main effects must be included before their corresponding interactions, meaning a model cannot incorporate an interaction term without the presence of the associated main effects \citep{bien_lasso_2013}. This requirement, known as the strong hierarchy property, is crucial for ensuring the meaningful interpretation of odds ratios within each cluster.
However, the conventional LASSO penalty treats all parameters independently, disregarding hierarchical relationships. As a consequence, it may shrink a main effect to zero while retaining its interaction term, leading to an incoherent model structure. To address this limitation, an alternative regularization approach, known as group-LASSO, has been proposed. Unlike standard LASSO, group-LASSO applies structured penalization, enforcing simultaneous shrinkage of predefined groups of variables. This ensures that any reduced model maintains strong hierarchy constraints, preserving both interpretability and consistency \citep{yuan_model_2006}.

To satisfy the strong hierarchy property, we adapt the approach proposed in \cite{lim_learning_2015}. For simplicity, we assume without loss of generality that $p = 1$, with $\pmb{X}_1$ being categorical with $L_1$ levels. The case $p > 1$ is a straightforward extension obtained by adding the appropriate terms to the expressions valid for $p = 1$. Under this assumption, we replace the linear predictor in the log-likelihood given by \eqref{eq:logisticstratalik} with the following expression:
\begin{equation}\label{eq:compositelinearpred}
	\beta_0 \mathbf{1} + \pmb{X}_1 \pmb{\alpha}_1 + \pmb{C} \pmb{\alpha}_\gamma + [\pmb{X}_1 \;\vdots\; \pmb{C} \;\vdots\; \pmb{\Xi}_1]
	\begin{pmatrix}
		\pmb{\alpha'}_1 \\
		\pmb{\alpha'}_{\pmb{\gamma}} \\
		\pmb{\alpha}_{\pmb{\theta}_1}
	\end{pmatrix},
\end{equation}
where the matrices $\pmb{X}_1$ and $\pmb{C}$ appear twice. Consequently, it is straightforward to derive the following relationships between the coefficients in \eqref{eq:compositelinearpred} and the actual parameters (with additions being component-wise):
\begin{eqnarray}
	\pmb{\beta}_1 & = & \pmb{\alpha}_1 + \pmb{\alpha'}_1,   \label{eq:component1} \\
	\pmb{\gamma} & = & \pmb{\alpha}_{\pmb{\gamma}} + \pmb{\alpha'}_{\pmb{\gamma}},  \label{eq:component2} \\
	\pmb{\theta}_1 & = & \pmb{\alpha}_{\pmb{\theta}_1}. \label{eq:component3}
\end{eqnarray}

The penalty is defined as:
\begin{equation}\label{eq:overlapgrouppenalty}
	\lambda \left( \lVert \pmb{\alpha}_1 \rVert_2 + \lVert \pmb{\alpha}_{\pmb{\gamma}} \rVert_2 + \sqrt{k \lVert \pmb{\alpha}_1' \rVert_2^2 + L_1 \lVert \pmb{\alpha}_{\pmb{\gamma}}' \rVert_2^2 + \lVert \pmb{\alpha}_{\pmb{\theta}_1} \rVert_2^2} \right),
\end{equation}
where $\lVert \cdot \rVert_2$ denotes the standard Euclidean $L_2$-norm. This penalty represents a specialized version of the overlapping group-LASSO \citep{hastie_statistical_2015}, as it applies an overlapping penalty to the parameters $(\pmb\beta_1^\top, \pmb\gamma^\top)^\top$. Depending on $\lambda \geq 0$, the term under the square root in \eqref{eq:overlapgrouppenalty} is zero if and only if:
\begin{equation}\label{eq:alphavectors}
	\lVert \pmb{\alpha}_1' \rVert_2^2 = \lVert \pmb{\alpha}_{\pmb{\gamma}}' \rVert_2^2 = \lVert \pmb{\alpha}_{\pmb{\theta}_1} \rVert_2^2 = 0,
\end{equation}
which occurs solely if each component of $(\pmb{\alpha}_1'^\top, \pmb{\alpha}_{\pmb{\gamma}}'^\top, \pmb{\alpha}_{\pmb{\theta}_1}^\top)^\top$ is the zero vector. In this case, the interaction terms are eliminated from the model. Apart from pathological solutions, the radius of the overlapping group-Lasso ball can only be small if all the vectors in \eqref{eq:alphavectors} are small, implying that they must be either all zero or all nonzero \citep{lim_learning_2015}. In other words, when interactions are present, they always appear alongside both main effects. This ensures that the simplified model automatically satisfies the strong hierarchy property. Cases in which interactions are included without one or both main effects are excluded due to the special geometry imposed by the optimization problem. 

Of course, in the case of continuous variables, since $\pmb{\alpha}_1$ reduces to a scalar, it follows that $\lVert \pmb{\alpha}_1 \rVert_2 = \lvert \alpha_1 \rvert$, thereby recovering the term that appears in the standard LASSO penalty. When multiple input variables are present, expression \eqref{eq:compositelinearpred} can be naturally extended--see, for instance, \cite{bilancia_interpretable_2024}--and the penalty \eqref{eq:overlapgrouppenalty} is adjusted accordingly. More importantly, \cite{lim_learning_2015} show that the solution to the overlapping group-LASSO problem is equivalent to that of a standard group-LASSO problem for a linear predictor expressed in terms of the actual parameters. Specifically, the penalties in both formulations are equivalent, and their corresponding linear predictors span the same column space. In this context, \cite{yuan_model_2006} demonstrate that the specific $L_2$-penalty used in the standard group-LASSO framework inherently imposes sum-to-zero constraints on the parameter estimates of categorical variables, thus resolving the identifiability issues associated with one-hot encoding. However, this alternative parameterization introduces additional complications, as we explain in greater detail in the following Section.

In principle, once the variables and interaction terms included in the sparse model have been identified, the model could be re-estimated with the same set of terms using the standard \texttt{glm()} function in \textsf{R} \citep{r_core_team_r_2025}, and the resulting coefficients interpreted accordingly. However, statistical inference following LASSO-based variable selection falls outside the scope of classical frequentist theory, as the set of selected active variables constitutes a random component of the model \cite{kammer_evaluating_2022}. Consequently, the associated confidence intervals are not valid, often yielding overly optimistic conclusions and compromising replicability. A more effective approach is to operate directly within the original parameterization of the sparse model, which is based on the sum-to-zero constraint. This formulation enables the seamless integration of results from simulation-based inference methods, such as bootstrap procedures, without the need to re-estimate a new logistic regression that includes only the variables and interactions active in the current bootstrap sample.

\subsubsection{Odds ratio calculation}\label{subsec:oddscalculation}

If $X_j$ is a continuous variable, the odds ratio is computed in the standard way by exponentiating the estimated regression coefficient. In the absence of an interaction term, the two odds ratios \eqref{eq:outsideOR} and \eqref{eq:insideOR} are identical, and the corresponding ratio of odds ratios equals 1.

When $X_j$ is a categorical variable with two levels and no interaction term, the same principle applies; however, the computation is complicated by the sum-to-zero constraint. The associated design submatrix $\pmb X_j$ comprises two columns, each corresponding to elements of the canonical basis of $\mathbb R^2$. The sum-to-zero constraint restricts the parameter estimates to the one-dimensional subspace $\pmb\beta_j = (-\beta_j, \beta_j)^\top$, implying that we can transform $X_j \in \{0, 1\}$ into $F(X_j) \in \{-1, +1\}$, where the first element of $\pmb\beta_j$ corresponds to $F(X_j) = -1$ and the second to $F(X_j) = 1$. Under this encoding, standard results for computing the adjusted odds ratio of a given input variable yield \citep{hosmer_applied_2013}:
\begin{align}
	& \log \mathsf{OR}(X_j = 1 \;\mathrm{ vs. }\; X_j = 0 \vert \textrm{rest}) \nonumber \\
	= &\; \mathrm{logit}(F(X_j) = 1 \vert \textrm{rest}) - \mathrm{logit}(F(X_j) = -1 \vert \textrm{rest})  \nonumber \\
	= &\; \beta_0 + (F(X_j) = 1) \beta_j  + \textrm{rest}  
	- \left[ \beta_0 + (F(X_j) = -1) \beta_j + \textrm{rest} \right]  \nonumber \\
	= &\; 2\beta_j,
\end{align}
which corresponds to the following rather unusual expression:
\begin{equation}\label{eq:ortwo_nointeractions}
	\mathsf{OR}(X_j = 1 \;\mathrm{ vs. }\; X_j = 0 \vert \textrm{rest}) = e^{2\beta_j},
\end{equation}
We emphasize that, in the conditional part, we have not specified the cluster to which we refer, since the interaction terms are null and, consequently, the odds ratio we have defined does not vary across clusters.

If $X_j \in \{0,1,2\}$ has three levels, in this case $\pmb X_j$ consists of three columns, and the estimates lie in the two-dimensional subspace defined by $\pmb \beta_j = (-\beta_{j1} - \beta_{j2}, \beta_{j1}, \beta_{j2})^\top$. This sum-to-zero parameterization is equivalent to express the variable $X_j$ into the following transformed form:
\begin{equation}\label{eq:sumtozeroX}
	\begin{matrix}
	&                 & \phantom{-}F_1{(X_j)}    & \phantom{-}F_2{(X_j)}  \\
	X_j = 0      & \longrightarrow & -1                & -1              \\
	X_j = 1      & \longrightarrow & +1      &  \phantom{-}0   \\
	X_j = 2      & \longrightarrow & \phantom{-}0      &  +1   \\   
	\end{matrix} 
\end{equation}
where, without loss of generality, the reference level is set to $X_j = 0$. Therefore:
\begin{align}\label{eq:exampleLogit}
	&\log \mathsf{OR}(X_j = 1 \;\mathrm{ vs. }\; X_j = 0 \vert \textrm{rest}) \nonumber \\
	= &\; \mathrm{logit}(F_1(X_j) = 1, F_2(X_j) = 0 \vert \textrm{rest}) -  \mathrm{logit}(F_1(X_j) = -1, F_2(X_j) = - 1 \vert \textrm{rest}) \nonumber	\\
	= &\; \beta_0 + (F_1(X_j) = 1)\beta_{j1} + (F_1(X_j) = 0)\beta_{j2} + \textrm{rest} \nonumber \\
	- &\: \left[ \beta_0 +  (F_1(X_j) = - 1)\beta_{j1} + (F_1(X_j) = - 1)\beta_{j2}\right] + \textrm{rest},
\end{align}
yielding:
\begin{equation}\label{eq:interactionOR1}
	\mathsf{OR}(X_j = 1 \;\mathrm{ vs. }\; X_j = 0 \vert \textrm{rest}) = e^{2\beta_{j1} + \beta_{j2}},
\end{equation}
and changing what needs to be changed:
\begin{equation}\label{eq:interactionOR2}
	\mathsf{OR}(X_j = 2 \;\mathrm{ vs. }\; X_j = 0 \vert \textrm{rest}) = e^{2\beta_{j2} + \beta_{j1}}.
\end{equation}

When the interaction term corresponding to a certain main effect is non-zero, we distinguish between different cases based on the number of elements in the data partition identified.

\paragraph{$\mathbf{k = 2}.$} 
When the identified partition consists of $k=2$ subsets, the matrix $\pmb C$ has two columns. For variables in which the interaction term is inactive, the procedure follows the approach previously outlined. In contrast, if $X_j$ is a categorical variable with two levels and the interaction term is active, the matrix $\pmb\Xi_j$ will have four columns, with the rows corresponding to the canonical basis of $\mathbb R^4$.
To better understand the nature of the parameter constraint at the level of the model's variables, we again set $F(X_j) \in \{-1, +1\}$. To maintain consistency also at the level of the variable indicating the cluster, in this section it will be denoted as $\mathcal C \in \{0,1\}$, with the obvious convention that $\mathcal C = 1$ implies $\mathcal D_1 = 1$ and $\mathcal D_0 = 0$, and similarly $\mathcal C = 0$ is equivalent to $\mathcal D_0 = 1$ and $\mathcal D_1 = 0$. At variance with the notation used in Subsection \ref{subsec:parameter_interpretation} based on the dummy coding, here the variables $(\mathcal D_0, \mathcal D_1)$ represent a one-hot encoding, and thus our convention is fully justified. Under the sum-to-zero parameterization, $F(\mathcal C) \in \{-1, +1\}$ thereby resulting in the following four possible combinations:
\begin{equation*}
\begin{array}{ccc}
	 & F(\mathcal C) = - 1 & F(\mathcal C) = + 1 \\
	F(X_j) = -1  & \theta_{j1} & -\theta_{j1}\\ 
	F(X_j) = +1  & -\theta_{j1} & \theta_{j1}
\end{array},
\end{equation*}
which clearly implies that the interaction parameter associated with the matrix $\pmb\Xi_j$ lies in a one-dimensional subspace defined by $(\theta_{j1}, -\theta_{j1})^\top$ when $F(\mathcal C) = -1$, and by $(-\theta_{j1}, +\theta_{j1})^\top$ when $F(\mathcal C) = +1$. By applying the same computations used in \eqref{eq:exampleLogit}, we immediately obtain that \eqref{eq:outsideOR} and \eqref{eq:insideOR} can be written, respectively, as:
\begin{align}
	\mathsf{OR}(X_j = 1 \;\mathrm{ vs. }\; X_j = 0\vert \mathcal D_1 = 1, \mathcal D_0 = 0, \textrm{rest}) & = e^{2(\beta_j + \theta_{j1})}, \\
	\mathsf{OR}(X_j = 1 \;\mathrm{ vs. }\; X_j = 0\vert \mathcal D_1 = 0, \mathcal D_0 = 1, \textrm{rest}) & = e^{2(\beta_j - \theta_{j1})},
\end{align} 
and these two expressions illustrate how \eqref{eq:ortwo_nointeractions} corresponds to the special case in which $\theta_{j1} = 0$. It then readily follows that $\textsf{ROR}(X_j) = \exp(4\theta_{j1})$.

If $X_j$ has three levels, then $\pmb \Xi_j$ has six columns, and we can consider the following matrix:
\begin{equation*}
\begin{array}{llc}
	 & & F(\mathcal C) = + 1  \\
	F_1(X_j) = -1, & F_2(X_j) = -1  & -\theta_{j1} - \theta_{j2} \\ 
	F_1(X_j) = +1, & F_2(X_j) = 0   & \theta_{j1} \\
	F_1(X_j) = 0, & F_2(X_j) = +1   & \theta_{j2} \\
\end{array},
\end{equation*}
since the terms belong to an additive linear predictor, and it is therefore appropriate to sum the corresponding effects with the correct sign. Accordingly, when $F(\mathcal C) = +1$, the resulting 2-dimensional subspace is $(-\theta_{j1} - \theta_{j2}, \theta_{j1}, \theta_{j2})^\top$. Similarly, it is straightforward to see that, with the necessary adjustments, for $F(\mathcal C) = -1$, the 2-dimensional subspace becomes $(\theta_{j1} + \theta_{j2}, -\theta_{j1}, -\theta_{j2})^\top$. In this case as well, it is intuitive to understand that the expressions of the odds ratios under this parameterization take the form:

\begin{align}\label{eq:threellevescov1}
	\mathsf{OR}(X_j = 1 \;\mathrm{ vs. }\; X_j = 0\vert \mathcal D_1 = 1, \mathcal D_0 = 0, \textrm{rest}) & = e^{2(\beta_{j1} + \theta_{j1}) + \beta_{j2} + \theta_{j2}},\\
	\mathsf{OR}(X_j = 2 \;\mathrm{ vs. }\; X_j = 0\vert \mathcal D_1 = 1, \mathcal D_0 = 0, \textrm{rest}) & = e^{2(\beta_{j2} + \theta_{j2}) + \beta_{j1} + \theta_{j1}},
\end{align} 
and:
\begin{align}\label{eq:threellevescov2}
	\mathsf{OR}(X_j = 1 \;\mathrm{ vs. }\; X_j = 0\vert \mathcal D_1 = 0, \mathcal D_0 = 1, \textrm{rest}) & = e^{2(\beta_{j1} - \theta_{j1}) + \beta_{j2} - \theta_{j2}},\\
	\mathsf{OR}(X_j = 2 \;\mathrm{ vs. }\; X_j = 0\vert \mathcal D_1 = 0, \mathcal D_0 = 1, \textrm{rest}) & = e^{2(\beta_{j2} - \theta_{j2}) + \beta_{j1} - \theta_{j1}},
\end{align} 
which, as a special case, include \eqref{eq:interactionOR1} and \eqref{eq:interactionOR2} when the interactions are null. As an immediate consequence:
\begin{align}
	\textsf{ROR}(X_j = 1) & = e^{4\theta_{j1} + 2\theta_{j2}}, \\
	\textsf{ROR}(X_j = 2) & = e^{4\theta_{j2} + 2\theta_{j1}}.
\end{align}

When $X_j$ is a continuous variable involved in non-zero interaction terms, its main effect is not subject to sum-to-zero coding. As an immediate consequence, direct calculations based on a unit increase in $X_j$ when evaluating the logits show that the correct expressions to be used are:
\begin{align}
	\mathsf{OR}(X_j = 1 \;\mathrm{vs.}\; X_j = 0\vert \mathcal D_1 = 1, \mathcal D_0 = 0, \textrm{rest}) & = e^{(\beta_j + \theta_{j1})}, \\
	\mathsf{OR}(X_j = 1 \;\mathrm{vs.}\; X_j = 0\vert \mathcal D_1 = 0, \mathcal D_0 = 1, \textrm{rest}) & = e^{(\beta_j - \theta_{j1})},
\end{align}
and thus $\textsf{ROR}(X_j) = \exp(2\theta_{j1})$.  
As in the previous case, when the interaction term is null, a single odds ratio--common to both clusters--applies, and is computed in the standard way as described at the beginning of Subsection \ref{subsec:oddscalculation}.

\paragraph{$\mathbf{k = 3}.$}

Using the same convention adopted for $k = 2$, we define in this case $\mathcal C \in \{0,1,2\}$ as the cluster indicator, where $\mathcal C = 0$ corresponds to $\mathcal D_0 = 1$, $\mathcal D_1 = 0$, and $\mathcal D_2 = 0$. Analogously, the remaining two values of $\mathcal C$ are associated with the corresponding elements of the canonical basis in $\mathbb R^3$. Under the sum-to-zero parameterization, the transformation applied to $\mathcal C$ coincides with the one defined in \eqref{eq:sumtozeroX} for $X_j$. When $X_j$ is a categorical variable with two levels, the matrix $\pmb\Xi_j$ contains six columns, while the associated parameters reside in a 2-dimensional subspace. A straightforward calculation shows that, for $\mathcal C=1$:
\begin{align}
	\mathrm{logit}(F(X_j) = 1 \vert F_1(\mathcal C) = 1, F_2(\mathcal C) = 0, \textrm{rest})  & =  \beta_0 + \beta_j + \gamma_1 + \theta_{j1}^{(1)} + \textrm{rest} \\
	\mathrm{logit}(F(X_j) = - 1 \vert F_1(\mathcal C) = 1, F_2(\mathcal C) = 0, \textrm{rest})  & =  \beta_0 - \beta_j + \gamma_1 - \theta_{j1}^{(1)} + \textrm{rest},
\end{align}
which implies that:
$$
\mathsf{OR}(X_j = 1 \;\mathrm{ vs. }\; X_j = 0\vert \mathcal D_1 = 1, \mathcal D_0 = 0, \mathcal D_2 = 0, \textrm{rest}) = e^{2(\beta_j + \theta_{j1}^{(1)})}
$$

However, when $\mathcal C = 0$ then $F_1(\mathcal C) = F_2(\mathcal C) = -1$, and it is immediate that:
\begin{align}
	\mathrm{logit}(F(X_j) = 1\vert F_1(\mathcal C) = -1, F_2(\mathcal C) =-1, \textrm{rest})  & =  \beta_0 + \beta_j - \gamma_1  - \gamma_2 - \theta_{j1}^{(1)} -\theta_{j1}^{(2)} + \textrm{rest} \label{eq:threelevels3}\\
	\mathrm{logit}(F(X_j) = - 1 \vert \vert F_1(\mathcal C) = -1, F_2(\mathcal C) = -1, \textrm{rest})  & =  \beta_0 - \beta_j - \gamma_1 - \gamma_2 + \theta_{j1}^{(1)} + \theta_{j1}^{(2)} + \textrm{rest}\label{eq:threelevels4},
\end{align}
paying close attention to the notational difference with \eqref{eq:threellevescov1} and \eqref{eq:threellevescov2}, where $\theta_{j1}$ and $\theta_{j2}$ denote the interaction parameters associated with the two levels of $X_j$ distinct from the reference category, while in the present context $\theta_{j1}^{(1)}$ and $\theta_{j1}^{(2)}$ represent the interaction parameters of $X_j$ (which has two levels) with the levels of $\mathcal C$ other than the one selected as the reference level.

Elementary algebraic manipulations reveal that \eqref{eq:threelevels3} and \eqref{eq:threelevels4}, when considered jointly, imply:
\begin{equation}\label{eq:rorthree1}
	\textsf{ROR}_{\mathcal C = 1}(X_j) = e^{4\theta_{j1}^{(1)} + 2\theta_{j1}^{(2)}},	
\end{equation}
where the notation $\textsf{ROR}_{\mathcal C=1}$ is used to highlight that the comparison concerns the strength of association between $X_j$ and the response in $\mathcal C_1$, relative to the corresponding estimate in $\mathcal C_0$. A similarly structured, symmetric expression holds for the remaining level:
\begin{equation}\label{eq:rorthree2}
	\textsf{ROR}_{\mathcal C = 2}(X_j) = e^{4\theta_{j1}^{(2)} + 2\theta_{j1}^{(1)}}.	
\end{equation}

As before, in the continuous case, computations can be carried out by considering a unit increase in $X_j$ when evaluating the logits--that is, by comparing $X_j = 1$ to $X_j = 0$. By simplifying the terms corresponding to $X_j = 0$, straightforward algebra then yields:
\begin{align}
	\textsf{ROR}_{\mathcal C=1}(X_j) & = e^{2\theta_{j1}^{(1)}+ \theta_{j2}^{(2)}}, \\
	\textsf{ROR}_{\mathcal C=2}(X_j) & = e^{2\theta_{j2}^{(2)}+ \theta_{j1}^{(1)}}.
\end{align}

When $X_j \in \{0,1,2\}$ has three levels, the matrix $\pmb \Xi$ has nine columns, and there are four free parameters, namely $(\theta_{j1}^{(1)}, \theta_{j2}^{(1)}, \theta_{j1}^{(2)}, \theta_{j2}^{(2)})^\top$, in contrast to the case in which $X_j$ has two levels, where only a single interaction parameter is associated with the cluster not chosen as the reference. The following expressions can be directly obtained by examining the logits involved in the computation of the odds ratios:
\begin{align}
	\textsf{ROR}_{\mathcal C=1} (X_j = 1) & = e^{4\theta_{j1}^{(1)} + 2\theta_{j2}^{(1)} + 2\theta_{j1}^{(2)} + \theta_{j2}^{(2)}}, \\
	\textsf{ROR}_{\mathcal C=1} (X_j = 2) & = e^{4\theta_{j2}^{(1)} + 2\theta_{j1}^{(1)} + 2\theta_{j2}^{(2)} + \theta_{j1}^{(2)}},
\end{align}

It is immediate to conclude that $\textsf{ROR}_{\mathcal C=1} (X_j=1)$ and $\textsf{ROR}_{\mathcal C=1}(X_j = 2)$ can be directly derived from one another by exchanging the indices in the subscript that denote the variables. Similarly, the ratios of odds ratios referring to $\mathcal C_2$ can be obtained by interchanging the superscripts that identify the cluster labels. For instance:
\begin{equation}
	\textsf{ROR}_{\mathcal C=2} (X_j = 1)= e^{4\theta_{j1}^{(2)} + 2\theta_{j2}^{(2)} + 2\theta_{j1}^{(1)} + \theta_{j2}^{(1)}}.
\end{equation}

\paragraph{$\mathbf{k > 3}.$} When the identified partition consists of more than three subsets, the symmetries observed in the previously derived expressions for the ratios of odds ratios can serve as a useful guide to obtain the desired results, without the need for direct computation based on logit comparisons. As an illustrative example, when $k = 4$ and $X_j \in \{0,1\}$ has two levels, it is immediate that three ratios of odds ratios will arise, and that \eqref{eq:rorthree1} and \eqref{eq:rorthree2} naturally extend as follows: 
\begin{align}
	\textsf{ROR}_{\mathcal C = 1}(X_j) & = e^{4\theta_{j1}^{(1)} + 2\theta_{j1}^{(2)} + 2\theta_{j1}^{(3)}}, \\	
	\textsf{ROR}_{\mathcal C = 2}(X_j) & = e^{4\theta_{j1}^{(2)} + 2\theta_{j1}^{(1)} + 2\theta_{j1}^{(3)}}, \\
	\textsf{ROR}_{\mathcal C = 3}(X_j) & = e^{4\theta_{j1}^{(3)} + 2\theta_{j1}^{(1)} + 2\theta_{j1}^{(2)}}.	
\end{align}
\newpage 
\section{Bootstrap inference}\label{sec:bootstrap}

As is customary, the regularization parameter $\lambda$ in \eqref{eq:overlapgrouppenalty} is selected via 10-fold cross-validation over a dense grid of values $\{\lambda_\ell\}_{\ell=1}^L$, in accordance with standard practice \citep{hastie_statistical_2015}. This procedure yields an average prediction error curve (averaged across folds) over the chosen grid. We then select the value $\hat\lambda_{\textrm{CV}}$ that minimizes this curve, which corresponds to the sparse coefficient estimate $\hat{\pmb\psi}(\hat\lambda_{\textrm{CV}})$. This notation emphasizes that the sparse parameter estimates depend on the estimated overlapping group penalty. When $\widehat\lambda_{\textrm{CV}}$ is large, few (if any) coefficients are shrunk to zero. Conversely, for small values of $\hat\lambda_{\textrm{CV}}$, the model tends to exhibit extreme sparsity, with most coefficients being driven toward zero \citep{hastie_09}.

The most straightforward approach to obtain the sampling distribution of $\hat{\pmb\psi}(\hat\lambda_{\textrm{CV}})$ while avoiding issues related to post-selection inference is to employ a non-parametric bootstrap \citep{Efron}. This involves drawing $B$ samples with replacement from the data $\{(y_i, \pmb x_i^\top, \mathcal c_i)\}_{i=1}^{N}$, where $\mathcal c_i$ denotes a generic indicator of the label assigned to the $i$-th student in the training set by the clustering algorithm. For each bootstrap sample $b$, with $b=1,\ldots,B$, we repeat the cross-validation procedure to identify the optimal $\hat\lambda_{\textrm{CV}}^{(b)}$ and the corresponding sparse estimate $\widehat{\pmb\psi}{}^{\ast}(\hat\lambda_{\textrm{CV}}^{(b)})$.  

To determine a final model, we begin with the straightforward observation that, for a given bootstrap sample, not all main effects and interactions will necessarily be active; in such cases, their corresponding coefficients are automatically set to zero. We then compute the proportion of times each coefficient is estimated to be zero across the bootstrap distribution and retain only those variables that meet a stringent inclusion criterion (for example, those whose estimated coefficients are zero in fewer than 10\% of the bootstrap samples).
Alternatively, since all relevant quantities are expressed as odds ratios, confidence intervals can be constructed using the $B$ bootstrap estimates available for each parameter \citep{justus2024}, and variables whose confidence intervals contain 1 may be excluded. In both approaches, the principle of strong hierarchy is automatically preserved. An illustration of these procedures is provided in Subsection \ref{subsec:assessing}.
Finally, we note the existence of a subtle limitation inherent in these procedures, stemming from the nature of the input variables; this issue will be further examined in Section \ref{sec:discussion}.

\section{Case study}\label{sec:example}
\subsection{Data}

The database used in this example is part of a larger dataset of students enrolled at the University of Foggia. During the pre-processing phase, all instances with missing data in at least one of the feature variables used in the analysis were removed. In total, we obtained a dataset of $n = 7595$ students enrolled between 2017 and 2023. Conventionally, for each academic year, the enrollment date was set to September 1st. Generally, enrollment opens on August 1st, although the vast majority of enrollments are actually finalized starting in September. 
Of these students, $n_G = 5618$ graduated (see Table \ref{tab:tab1}), while $n_D = 1971$ experienced early dropout. In this study, early dropout is defined as follows: if enrollment occurred on 09/01/\verb+yyyy+, where \verb+yyyy+ represents the enrollment year, dropout was finalized within the following 16 months, that is, by 12/31/\verb+yyyy++1 at the latest.  
Notably, the database also includes graduates from the 2023 enrollment cohort. This is possible because, under the Italian university system, exams taken during previous academic careers can be recognized as valid, significantly shortening the study path.

The other variables listed in Table \ref{tab:tab1}, whose meaning is not immediately clear, are as follows:
\begin{itemize}
	\item[-] High-school final grade: In Italy, the high school diploma is obtained through a state exam, with a minimum score of 60 and a maximum score of 100. In the latter case, an honors distinction may be awarded, but this possibility has been ignored in order to treat the variable as a numerical one.
	\item[-] Late enrollment: If \verb+yyyy+ denotes the year in which the high-school diploma is obtained, enrollment is considered late if the enrollment year is greater than or equal to \verb+yyyy++1. This can occur for various reasons, such as enrolling in a university degree program at an older age, often while concurrently working.
	\item[-] Degree: The university reform implemented in Italy in 1999 subdivided university education into two main cycles: the Bachelor's degree (or first-level degree) and the Master's degree (or second-level degree). Single-cycle degree programs confer a Master's degree after a continuous cycle of 5 or 6 years in various disciplines, which are regulated by specific European protocols (e.g., Medicine, Veterinary Medicine, Law, Architecture, etc.).
	\item[-] ISEUU: It is a value that certifies the economic situation of the student's family to determine eligibility for exemption from tuition fees and enrollment in a degree program. It can be considered a proxy for the socio-economic condition of the family in which the student resides.
\end{itemize}

	\begin{table}[!htb] \centering \renewcommand*{\arraystretch}{1.1}\caption{\textit{Summary statistics for $n = 7595$ students, subdivided into graduates and early dropouts.}}\label{tab:tab1}\resizebox{\textwidth}{!}{
		\begin{tabular}{lrrrrrrl}
			\hline
			\hline
			Outcome & \multicolumn{3}{c}{Early dropout} & \multicolumn{3}{c}{Graduated} &   \\ 
			Variable & \multicolumn{1}{c}{$N$} & \multicolumn{1}{c}{Mean} & \multicolumn{1}{c}{$SD$} & \multicolumn{1}{c}{$N$} & \multicolumn{1}{c}{Mean} & \multicolumn{1}{c}{$SD$} & \multicolumn{1}{c}{Test} \\ 
			\hline
			Sex & 1977 &  &  & 5618 &  &  & $\chi^2=27.317^{***}$ \\ 
			... F & 1224 & 62\% &  & 3842 & 68\% &  &  \\ 
			... M & 753 & 38\% &  & 1776 & 32\% &  &  \\ 
			High-school final grade  & 1977 & 81 & 12 & 5618 & 82 & 12 & $F=7.479^{***}$ \\ 
			Late enrollment & 1977 &  &  & 5618 &  &  & $\chi^2=319.705^{***}$ \\ 
			... No & 1258 & 64\% &  & 2263 & 40\% &  &  \\ 
			... Yes & 719 & 36\% &  & 3355 & 60\% &  &  \\ 
			Age at enrollment (yrs.)& 1977 & 21 & 5 & 5618 & 23 & 5 & $F=67.08^{***}$ \\ 
			Residence Region & 1977 &  &  & 5618 &  &  & $\chi^2=6.932^{***}$ \\ 
			... Apulia & 1829 & 93\% &  & 5085 & 91\% &  &  \\ 
			... Elsewhere & 148 & 7\% &  & 533 & 9\% &  &  \\ 
			Degree & 1977 &  &  & 5618 &  &  & $\chi^2=593.934^{***}$ \\ 
			... Bachelor's & 1677 & 85\% &  & 3564 & 63\% &  &  \\ 
			... Master's & 136 & 7\% &  & 1895 & 34\% &  &  \\ 
			... Single cycle & 164 & 8\% &  & 159 & 3\% &  &  \\ 
			ISEEU (\texteuro) & 1977 & 17534 & 13358 & 5618 & 18735 & 14763 & $F=10.172^{***}$\\ 
			\hline
			\hline
			\multicolumn{8}{l}{Statistical significance markers: *$p<0.1$; **$p<0.05$; ***$p<0.01$}\\ 
		\end{tabular}
	}
\end{table}

Table \ref{tab:tab1} presents the comparisons between the two groups for all included variables. The differences are significant in all cases, and in some instances, interesting patterns emerge. For example, students who enrolled late have a considerably higher graduation rate compared to those who enrolled in the same year as their high school graduation (60\% vs. 36\%), likely as a result of greater motivation stemming from a more conscious decision. Beyond these descriptive observations, we cannot draw further conclusions about the relative weight of the variables in explaining their impact on the final outcome of each student's academic career.

\subsection{Searching the optimal partition}\label{sec:searching}

In what follows we strictly adhere to the notation introduced in Section \ref{sec:notation}. To determine an optimal partition, we considered distance-based algorithms, as they are often less susceptible to tuning issues compared to their model-based counterparts \citep{preudhomme_head--head_2021}. This approach requires defining a dissimilarity matrix suitable for heterogeneous data that include both continuous and categorical variables. In such cases, the Gower similarity coefficient is commonly recommended \citep{gower_general_1971,dorazio_distances_2021}:  
\begin{equation}  
	S(\pmb d_i, \pmb d'_i) = \frac{1}{p+1}\left(\sum_{j=1}^q \left( 1 -\frac{\vert d_{ij} - d_{i'j} \vert }{R_j} \right) + \sum_{j=q+1}^{p+1} s(d_{ij}, d_{i'j})\right),  
\end{equation}  
where $s(d_{ij}, d_{i'j}) = 1$ if $d_{ij} = d_{i'j}$ and $0$ otherwise, while $R_j = \max(d_j) - \min(d_j)$. As implemented in the \texttt{daisy} function of the \texttt{cluster} \textsf{R} package \citep{r_core_team_r_2025,maechler_cluster_2024}, the Gower dissimilarity is obtained through the transformation $1 - S(\pmb d_i, \pmb d'_i)$.

With Gower dissimilarity, the Partitioning Around Medoids (PAM) algorithm is a well-suited non-hierarchical choice, as it is generally more robust than $k$-means, being less sensitive to outliers and the assumption of spherical clusters \citep{budiaji_simple_2019,schubert_fast_2021,botyarov_partitioning_2022,nouraei_comparison_2022}. Moreover, the resulting partitions can be readily used to assign new observations to the previously defined clusters. The algorithm iterates through the following phases until convergence is reached \citep{reynolds_clustering_2006}:
\begin{itemize}
	\item[-] BUILD: A collection of $k$ representative objects (medoids) is selected. The first medoid is chosen as the data point that minimizes the sum of distances (or energy) to all other data points. The second medoid is selected as the data point that results in the greatest reduction of the total distance between each data point and its nearest medoid. These steps are repeated until $k$ medoids have been selected.
	\item[-] SWAP: An attempt is made to refine the set of selected medoids, thereby improving the quality of clustering. Different data points are tested as potential medoids, and if replacing a medoid with another point decreases the energy, PAM performs the swap. All possible combinations are evaluated, ensuring that only one optimal solution is obtained.
\end{itemize}

To determine the optimal number of components, rather than relying solely on traditional approaches based on the geometric properties of clusters, we opted for a stability-based approach. This method evaluates the output of a clustering algorithm by assessing the similarity of partitions generated from data sets drawn from the same probabilistic source \citep{luxburg_clustering_2010,liu_stability_2022}. By circumventing the bias inherent in purely geometric criteria and instead focusing on the structure of the data-generating process, this approach is expected to yield subgroups that are often more interpretable from the perspective of applied researchers.

In particular, we followed the approach proposed by \cite{hennig_cluster-wise_2007} for non-parametric bootstrapping of the Jaccard similarity coefficient between the original partition and a perturbed version induced by bootstrap resampling. In summary, let $\pmb D_n^{(b)}= (\pmb d_1^{(b)}, \ldots, \pmb d_n^{(b)})^\top$ be the bootstrap sample of size $n$ obtained by resampling $\pmb D_n$ with replacement. Let $\pmb D_\star^{(b)}$ be the data matrix corresponding to the $b$-th bootstrap sample, from which all duplicate points have been removed, ensuring that each data point appears exactly once after this step (except in cases where repeated data points were already present in the original dataset). Let $\mathbb P^{(b)} =  C_k(\pmb D_\star^{(b)})$ denote the partition of size $k$ induced by the $b$-th bootstrap sample without duplicates. If $\mathcal C \in C_k(\pmb D_n)$ is a cluster of the original partition, let $\mathcal C_\star^{(b)}$ denote the restriction of $\mathcal C$ to the data points in $\pmb D_\star^{(b)}$.  
We assess the agreement between $\mathcal C$ and the partition induced by the $b$-th bootstrap sample using the Jaccard similarity between sets $A$ and $B$:  
\begin{equation}
	J(A, B) = \frac{\vert A \cap B \vert }{\vert A \cup B \vert },
\end{equation}  
which quantifies the percentage of agreements in co-membership. For each bootstrap sample, $b=1,\ldots,B$, if $\mathcal C_\star^{(b)}\neq \emptyset$, we define:  
\begin{equation}
	J_{\mathcal C, k}^{(b)} = \max_{S \in \mathbb P^{(b)}_k} J(\mathcal C_\star^{(b)}, S).
\end{equation}  

In \cite{hennig_cluster-wise_2007}, the variability across bootstrap samples is averaged out by computing the following overall similarity measure between the cluster $\mathcal C$ of the original partition and its perturbed bootstrap versions:
\begin{equation} J_{\mathcal C, k} = \frac{1}{B_{\mathrm{eff}}} \sum_{b=1}^B J_{\mathcal C, k}^{(b)} \end{equation}
where $B_{\mathrm{eff}}$ denotes the number of bootstrap samples for which $\mathcal C_\star^{(b)}\neq \emptyset$.
To define a criterion for selecting the optimal number of clusters, borrowing from \cite{yu_bootstrapping_2019}, we define the optimal value $k^\star$ as:
\begin{equation}\label{eq:finalcriterion} 
	k^\star = \underset{k = 2,\ldots,k_{\max}}{\mathrm{argmax}}\left\{\underset{\mathcal C \in C_k(\pmb D_n)}{\mathrm{min}} J_{\mathcal C,k} \right\}, 
\end{equation}
i.e., $k^\star$ is the value that maximizes the worst-case similarity for each $k$. In the presence of ties, we select the smallest value of $k^\star$ that satisfies \eqref{eq:finalcriterion}, following a principle of parsimony. 

\begin{figure*}[!h] 
	\begin{center}
		\includegraphics[width=0.49\linewidth]{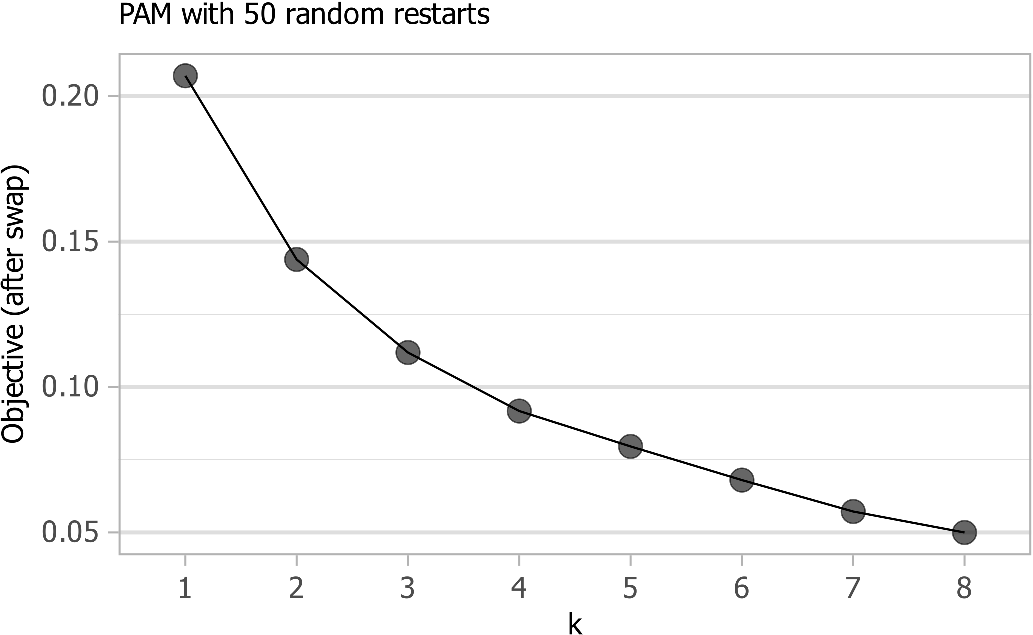}\hskip 0.3cm
		\includegraphics[width=0.49\linewidth]{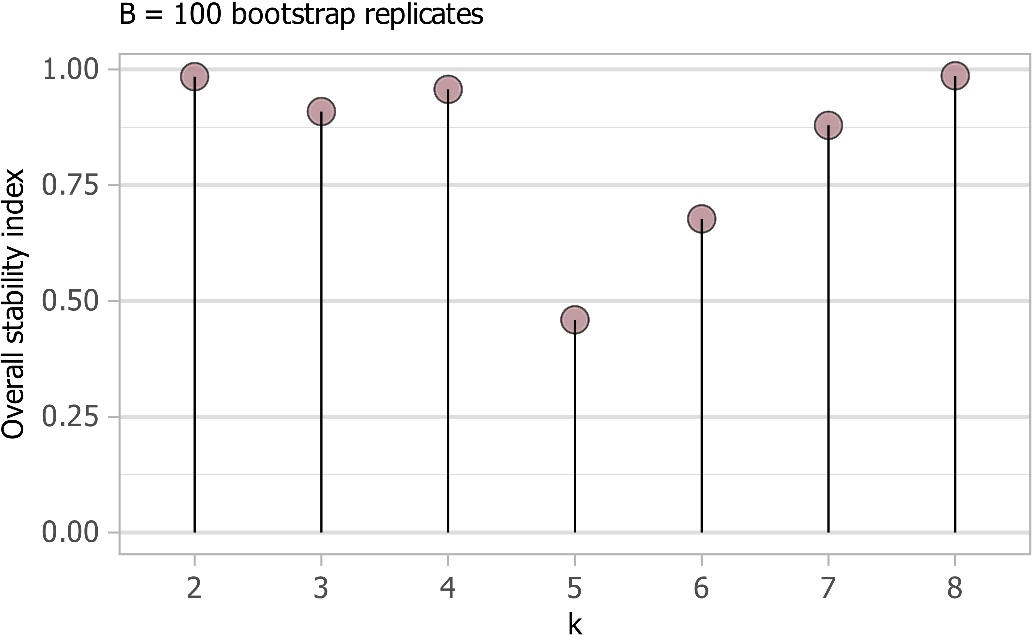}
	\end{center}     
	\caption{\textit{Left panel: Final energy value of the PAM algorithm after the swap phase, once convergence is achieved, for $k$ values ranging from 2 to 8, with 50 random restarts for each value of $k$. Right panel: Bootstrap stability criterion \eqref{eq:finalcriterion} based on $B = 100$ bootstrap replicates.}}\label{fig:partition}
\end{figure*}

The results are presented in Figure \ref{fig:partition}. The graph on the left displays the final energy value after the swap phase, once convergence is achieved, for $k$ values ranging from 2 to 8. Since each algorithm run explores a distinct local maximum of the surface defined by the objective function, for each $k$, the algorithm was executed 50 times from random initializations, with the run yielding the highest local maximum being selected. The graph on the right shows the value of the stability criterion \ref{eq:finalcriterion} based on $B = 100$ bootstrap replicates. In accordance with the principle of parsimony, the optimal value is naturally chosen as $k^\star = 2$, particularly considering that the left-hand graph reveals the steepest decrease in energy when transitioning from a complete absence of heterogeneity, represented by a single cluster, to the run with $k=2$ clusters.

Table \ref{tab:tab2} presents the comparisons between the clusters. Unlike Table \ref{tab:tab1}, the outcome variable is also included among the variables being compared, while the pivoting variable is the label of the clustering identified by the PAM algorithm at the optimal value $k^\star = 2$. As in the previous case, potentially interesting insights emerge, such as the fact that some variables are strongly separated between the two groups. For instance, 100\% of the students who enroll late are placed in cluster $\mathcal C_2$. However, even in this case, we are limited to making essentially descriptive observations.

\begin{table}[!htb] \centering \renewcommand*{\arraystretch}{1.1}\caption{\textit{Summary statistics for $n = 7595$ students, using the clustering label as the pivoting variable.}}\label{tab:tab2}\resizebox{\textwidth}{!}{
		\begin{tabular}{lrrrrrrl}
			\hline
			\hline
			Partition & \multicolumn{3}{c}{$\mathcal C_1$} & \multicolumn{3}{c}{$\mathcal C_2$} &   \\ 
			Variable & \multicolumn{1}{c}{$N$} & \multicolumn{1}{c}{Mean} & \multicolumn{1}{c}{$SD$} & \multicolumn{1}{c}{$N$} & \multicolumn{1}{c}{Mean} & \multicolumn{1}{c}{SD} & \multicolumn{1}{c}{Test} \\ 
			\hline
			Sex & 4311 &  &  & 3284 &  &  & $\chi^2=4.463^{**}$ \\ 
			... F & 2919 & 68\% &  & 2147 & 65\% &  &  \\ 
			... M & 1392 & 32\% &  & 1137 & 35\% &  &  \\ 
			High-school final grade & 4311 & 85 & 11 & 3284 & 77 & 11 & $F=887.858^{***}$ \\ 
			Late enrollment & 4311 &  &  & 3284 &  &  & $\chi^2=4990.473^{***}$ \\ 
			... No & 3520 & 82\% &  & 1 & 0\% &  &  \\ 
			... Yes & 791 & 18\% &  & 3283 & 100\% &  &  \\ 
			Age at enrollment & 4311 & 20 & 3 & 3284 & 25 & 6 & $F=2846.112^{***}$ \\ 
			Residence Region & 4311 &  &  & 3284 &  &  & $\chi^2=93.139^{***}$ \\ 
			... Apulia & 4044 & 94\% &  & 2870 & 87\% &  &  \\ 
			... Elsewhere & 267 & 6\% &  & 414 & 13\% &  &  \\ 
			Degree & 4311 &  &  & 3284 &  &  & $\chi^2=3670.555^{***}$ \\ 
			... Bachelor & 4109 & 95\% &  & 1132 & 34\% &  &  \\ 
			... Master & 0 & 0\% &  & 2031 & 62\% &  &  \\ 
			... Single cycle & 202 & 5\% &  & 121 & 4\% &  &  \\ 
			ISEEU & 4311 & 18036 & 14059 & 3284 & 18930 & 14865 & $F=7.162^{***}$ \\ 
			Outcome & 4311 &  &  & 3284 &  &  & $\chi^2=324.619^{***}$ \\ 
			... Dropout & 1464 & 34\% &  & 513 & 16\% &  &  \\ 
			... Graduated & 2847 & 66\% &  & 2771 & 84\% &  & \\ 
			\hline
			\hline
			\multicolumn{8}{l}{Statistical significance markers: *$p<0.1$; **$p<0.05$; ***$p<0.01$}\\ 
		\end{tabular}
	}
\end{table}

\subsection{Standard logistic regression with cluster-based interactions}\label{sec:standard}

The estimates presented in this Section are particularly instructive for the correct interpretation of the \textsf{ROR} measure defined in \eqref{eq:effectratio}. In Table \eqref{tab:tab3}, $\mathcal C_1$ is taken as the reference level of the partition indicator, while the calculations of \textsf{RORs} are performed for $\mathcal C_2$ vs.~$\mathcal C_1$. From Table \eqref{tab:tab3}, it is evident that the interaction term corresponding to the `Master' level of the variable \verb+degree+ is not estimable. This is a consequence of the sparsity observed in $\mathcal C_1$ for this variable (there are no students in $\mathcal C_1$ enrolled in a Master’s degree). Indeed, since the `Master' level is absent in $\mathcal C_1$, the variable encoding the indicator for $\mathcal C_2$ is identical to the dummy variable coding for this level, and consequently, the column encoding the interaction is also identical, making it impossible to estimate either of the two conditional odds ratios. Although the variable \texttt{late\_enrollment} exhibits the same issue, in this case, the cluster in which it occurs is $\mathcal C_2$, where all students have enrolled late. However, since $\mathcal C_2$ is not the reference level, the two columns of the model matrix corresponding to the main effect and the interaction will not, in general, be identical as in the previous case.  

Apart from pathological cases, when $\mathsf{ROR}(X_j) > 1$, we can assert that the association between $X_j$ and the outcome (graduation) is $100\times(\mathsf{ROR}(X_j) - 1)\%$ stronger within $\mathcal C_2$ than within $\mathcal C_1$. *Mutatis mutandis*, a similar interpretation applies when $\mathsf{ROR}(X_j) < 1$, with $100\times(\mathsf{ROR}(X_j) - 1)\%$ replaced by $100\times(1 - \mathsf{ROR}(X_j))\%$. For instance, the association between late enrollment and graduation is approximately 100\% stronger in cluster $\mathcal C_2$ than in cluster $\mathcal C_1$. Similarly, in cluster $\mathcal C_2$, where students graduate more frequently (84\% in $\mathcal C_2$ vs.~66\% in $\mathcal C_1$), the association between male gender and graduation is approximately 11\% stronger than in cluster $\mathcal C_1$, although in both groups male students are at a higher risk of dropping out (with both odds ratios being $< 1$). All these conclusions provide evidence that should be taken into account through targeted policy interventions.

To further clarify this point, it is important to recall that the model can also be employed in a predictive context. For instance, with the PAM algorithm, a future instance can be assigned to a specific partition, determined during the training phase, by allocating it to the nearest medoid center \citep{aggarwal13}. In this way, based on the parameter estimates obtained during training, we may expect that future students assigned to $\mathcal C_2$ should be monitored particularly closely when they do not enroll late. The increased likelihood of dropout associated with this scenario is plausibly a consequence of the fact that a student who enrolls late--whether because they had more time to evaluate the demands and requirements of a Bachelor's or Single Cycle program, or because they enroll in a Master's program after already completing a degree (which, by definition, makes them a late-enrolling student)--has undergone a more deliberate decision-making process. Similarly, male students assigned to $\mathcal C_1$ should receive increased attention, as they exhibit the highest risk of dropout. These findings make it possible to identify a limited number of critical factors on which to concentrate resources, with a level of granularity that would not be attainable using a standard logistic regression model.

Although this study primarily focuses on the in-sample characterization of student subpopulations and their heterogeneity concerning dropout propensity, the proposed model is also suitable for predictive applications. Specifically, it can be employed to forecast the outcome variable--dropout or graduation--for prospective students. We expect that this model will achieve superior predictive accuracy compared to standard logistic regression, owing to its capacity to capture interaction effects and latent group structures. However, a comprehensive empirical evaluation of its predictive performance remains to be conducted.

\begin{table}[!htb] \centering \renewcommand*{\arraystretch}{1.1}\caption{\textit{\textsf{Model matrix}: For the column $\textsf{ROR}$, this refers to the columns of the model matrix associated with the interaction terms between the variable of interest (dummy coded for categorical variables) and the dummy-coded indicator of the partition (with $\mathcal C_1$ as the reference level). In contrast, the columns $\textsf{OR}_{\texttt{glm}}\;\mathcal C_1$(a) and $\textsf{OR}_{\texttt{glm}}\;\mathcal C_2$(b) correspond to main effects.
			$\textsf{ROR}$: Ratio of odds ratios \eqref{eq:effectratio} comparing $\mathcal{C}_2$ to $\mathcal{C}_1$, estimated from the full logistic regression model \eqref{eq:logisticstratalik}, which includes interaction terms with the indicator variable for the identified partition.
			$\textsf{OR}_{\texttt{glm}}\;\mathcal C_1$(a): Odds ratios from a standard logistic regression model without interactions, estimated using only the observations in subset $\mathcal C_1$.
			$\textsf{OR}_{\texttt{glm}}\;\mathcal C_2$(b): Odds ratios from a standard logistic regression model without interactions, estimated using only the observations in subset $\mathcal C_2$.}}\label{tab:tab3}\resizebox{0.80\textwidth}{!}{
		\begin{tabular}{lcccr}
			\hline\hline
			\textsf{Model matrix} & \textsf{ROR} & $\textsf{OR}_{\texttt{glm}}\;\mathcal C_1$(a) & $\textsf{OR}_{\texttt{glm}}\;\mathcal C_2$(b) & (b)/(a)\\
			\hline
			\texttt{sex}(M) & 1.1173 & 0.7473 & 0.8350 & 1.1174\\
			\texttt{high\_school\_final\_grade} & 0.9968 & 1.0153 & 1.0121 & 0.9968\\
			\texttt{late\_enrollment}(Yes) & 0.0001 & 1.6021 & 0.0000 & 0.0000\\
			\texttt{age\_at\_enrollment} & 1.0201 & 0.9566 & 0.9758 & 1.0201\\
			\texttt{residence\_region}(Elsewhere) & 0.8228 & 1.3818 & 1.1370 & 0.8228\\
			\texttt{degree}(Master) & \texttt{NA} & \texttt{NA} & 4.9867 & \texttt{NA}\\
			\texttt{degree}(Single cycle) & 0.9099 & 0.4116 & 0.3745 & 0.9099\\
			\texttt{iseuu} & 1.0000 & 1.0000 & 1.0000 & 1.0000\\
			\hline
		\end{tabular}
	}
\end{table}

We also point out that in Table \eqref{tab:tab3}, we report the estimated Odds Ratios from a standard model without interaction terms, fitted separately on the subset of students assigned to cluster $\mathcal{C}_1$, see column $\textsf{OR}_{\texttt{glm}}\;\mathcal{C}_1$(a), with an analogous interpretation for column $\textsf{OR}_{\texttt{glm}}\;\mathcal{C}_2$(b). Therefore, the ratios between these Odds Ratios, as shown in column (b)/(a), provide a straightforward numerical confirmation--disregarding negligible discrepancies due to floating-point representation--of the ``outside vs.~inside'' interpretation of the two conditional odds ratios in \eqref{eq:outsideOR} and \eqref{eq:insideOR}, as discussed in Section \ref{subsec:parameter_interpretation}.

\subsection{Sparse logistic regression with interactions}

\begin{figure*}[!h] 
	\begin{center}
		\includegraphics[width=0.70\linewidth]{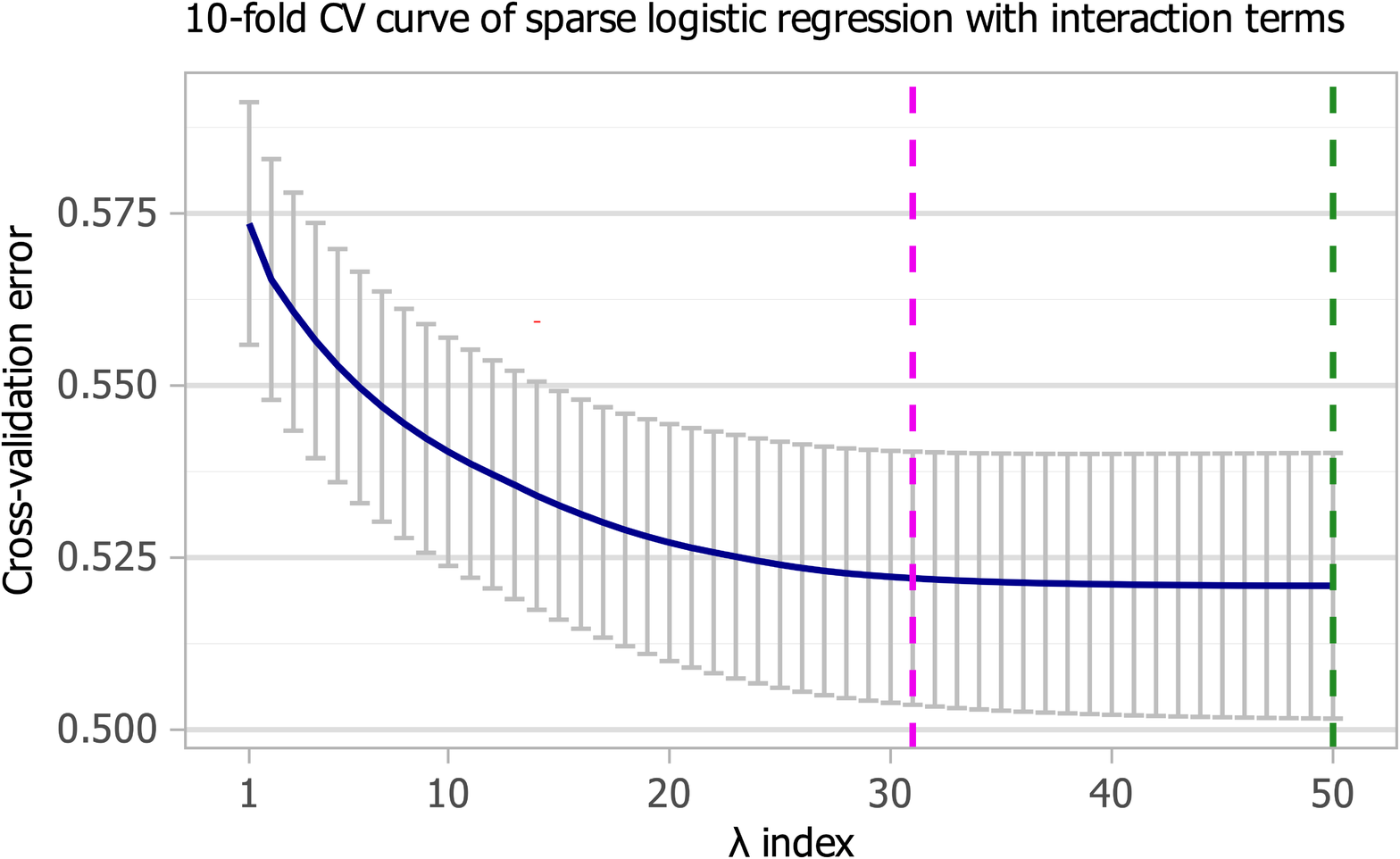}\hskip 0.3cm
	\end{center}     
	\caption{\textit{Blue curve: regularization path over a $\lambda$-grid, with the generalization error estimated via 10-fold cross-validation (approximate confidence intervals are superimposed). Green dashed vertical line: optimal value of $\lambda$, corresponding to the minimum generalization error. Purple dashed vertical line: value of $\lambda$ associated with a sub-optimal solution, defined as the largest value for which at least $p \leq 3$ interaction terms remain active in the model.}}\label{fig:sparse_CV}
\end{figure*}

We estimated a sparse logistic model in which the interactions adhere to the strong hierarchy property discussed in Subsection~\ref{sec:estimation}. The regularization parameter, $\lambda$, was selected via 10-fold cross-validation, in line with standard practice \citep{hastie_statistical_2015}. All computations were performed using \textsf{R}~4.4.2 and the \texttt{glInternet}~1.0.12 package \citep{r_core_team_r_2025,glinternet25}. Given that cross-validation inherently involves a random component due to the partitioning of the training dataset into folds, we repeated the execution of the group-LASSO algorithm 50 times to identify an optimal sparse solution, using the same $\lambda$-grid in each repetition. The blue curve in Figure~\ref{fig:sparse_CV} corresponds to the run that yielded the lowest estimate of the generalization error (cross-validation error) across all minimum values obtained in the 50 repetitions, thus reducing the dependency on fold assignment.

To enable comparison with the standard parameter estimates obtained via the \texttt{glm()} function, we temporarily imposed a sub-optimal solution using the same $\lambda$-grid previously employed. In Figure~\ref{fig:sparse_CV}, this particular solution corresponds to the value of $\lambda$ indicated by the intersection with the purple vertical dashed line. As is well known, when $\lambda$ is large, few (if any) coefficients are shrunk to zero. Conversely, for small values of $\lambda$, the model tends to become highly sparse, with most coefficients being driven towards zero \citep{hastie_09}. The sub-optimal solution considered here is based on a heuristic criterion: $\lambda$ is selected as the largest value for which at most $p \leq 3$ interaction terms remain active in the model.

In Table~\ref{tab:tab4} we have reported in detail the results obtained with both algorithmic solutions. The obtained results in the two cases are only slightly different, and they remain generally consistent in terms of both magnitude and direction (i.e., whether the estimated odds ratio is greater or less than 1). This discrepancy arises from the fact that the group-LASSO algorithm employs a fast variant of the Iterative Shrinkage-Thresholding Algorithm (ISTA), a first-order optimization method based on gradient descent. Specifically, it updates the coefficient vector iteratively by taking a step in the direction of the negative gradient of the least-squares component of the LASSO objective function, followed by a shrinkage/thresholding operation applied to each coefficient \citep{zhao2023surveynumericalalgorithmssolve}.

By way of example, in the sparse model, male students exhibit a higher risk of dropping out than female students in both clusters; in particular, this risk is approximately 6\% higher for male students assigned to cluster $\mathcal C_2$, while this difference increases to 13\% when calculated using the \texttt{glm()} function. For the remaining variables, concerns arise regarding the statistical significance of the corresponding odds ratios. For instance, the value associated with \texttt{high\_school\_final\_grade} corresponds to a \textsf{ROR} equal to 1 when rounded. However, in the absence of a formal assessment of statistical significance based on the procedures described in Section~\ref{sec:bootstrap}, and given the sub-optimal nature of the solution under consideration, we defer any substantive interpretation to the following paragraph. As previously mentioned, we cannot directly use $p$-values from the refitted model with the \texttt{glm()} function, due to the inconsistencies that affect these quantities when they are computed within an inference post-model selection framework.

\begin{table}[!ht] \centering \renewcommand*{\arraystretch}{1.1}\caption{\textit{\textsf{OR}(glm): estimated odds ratio computed from the main effects selected by the sub-optimal sparse model (with at most $p \leq 3$ active interaction terms), re-estimated using the standard \texttt{glm()} function and interpreted according to their conditional interpretation defined in Subsection \ref{subsec:parameter_interpretation}.
			\textsf{OR}(sparse): odds ratio computed from the main effects and interactions of the sub-optimal sparse model, corresponding to the largest value of $\lambda$ for which at most $p \leq 3$ interaction terms remain active, calculated directly from the expression reported in Subsection \ref{subsec:oddscalculation}, and interpreted according to their conditional interpretation defined in Subsection \ref{subsec:parameter_interpretation}.
			\textsf{ROR}(glm): ratio of odds ratios calculated from the estimates obtained with the \texttt{glm()} function, corresponding to the odds ratio estimated on the interaction terms according to \eqref{eq:effectratio}.
			\textsf{ROR}(sparse): ratio of odds ratios computed from the sub-optimal sparse model with at most $p \leq 3$ active interaction terms, using the variables associated with the active interactions (the corresponding main effects are highlighted in bold). For variables with inactive interaction terms, the outside and inside $\mathcal C_2$ odds ratios are identical, and the corresponding ratio of odds ratios equals 1. For simplicity, these trivial \textsf{ROR}s have been omitted from the table.}}\label{tab:tab4}\resizebox{0.80\textwidth}{!}{
		\begin{tabular}{lcccc}
			\hline\hline
			& \textsf{OR}(glm) & \textsf{OR}(sparse) & \textsf{ROR}(glm) & \textsf{ROR}(sparse)\\
			\hline
			       &  Outside $\mathcal C_2$ & Outside $\mathcal C_2$ & \\
			       \texttt{high\_school\_final\_grade} & 1.0151 & \textbf{1.0112} & -- & --\\
			       \texttt{age\_at\_enrollment} & 0.9710 & 0.9847 & -- & --\\
			       \texttt{iseeu} & 1.0000 & 1.0000 & -- & --\\
			\texttt{\textbf{sex}}(M) & 0.7443 &  \textbf{0.8078} & -- & --\\
			\texttt{late\_enrollment}(Yes) & 1.5207 & 1.4273 & -- & --\\
			\texttt{residence\_region}(Elsewhere) & 1.3731 & \textbf{1.1286} & -- & --\\
			\texttt{degree}{(Master)} & 5.1261 & 4.2300 & -- & --\\
			\texttt{degree}{(Single cycle)} & 0.4005 & 0.5182 & -- & --\\
			\texttt{cluster}($\mathcal C_2$) & 1.5714 & 1.1155 & -- & --\\
			\hline
			      &   & Inside $\mathcal C_2$  & \\
			\texttt{high\_school\_final\_grade}:\texttt{cluster}($\mathcal C_2$) & -- & 1.0111 & 0.9963 &  1.0000\\
			\texttt{sex}(M):\texttt{cluster}($\mathcal C_2$) & -- & 0.8603 & 1.1392 & 1.0650\\
			\texttt{residence\_region}(Elsewhere):\texttt{cluster}($\mathcal C_2$) & -- & 0.9913 & 0.8401 & 0.8783 \\
			\hline
		\end{tabular}
	}
\end{table}

\subsection{Assessing statistical significance}\label{subsec:assessing}

\begin{figure*}[!h] 
	\begin{center}
		\includegraphics[width=0.49\linewidth]{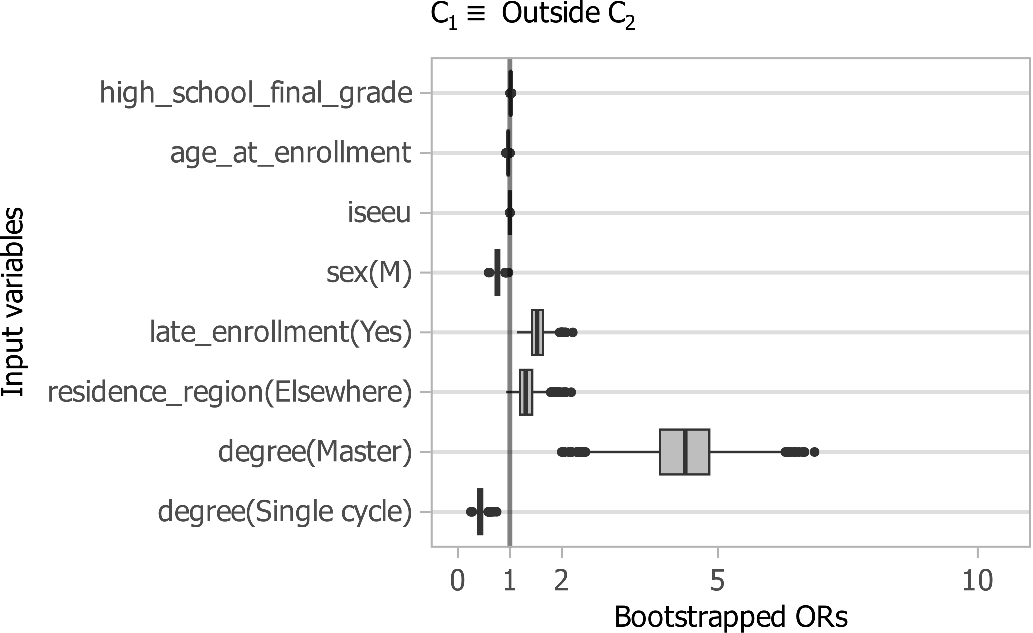}\hskip 0.3cm
		\includegraphics[width=0.49\linewidth]{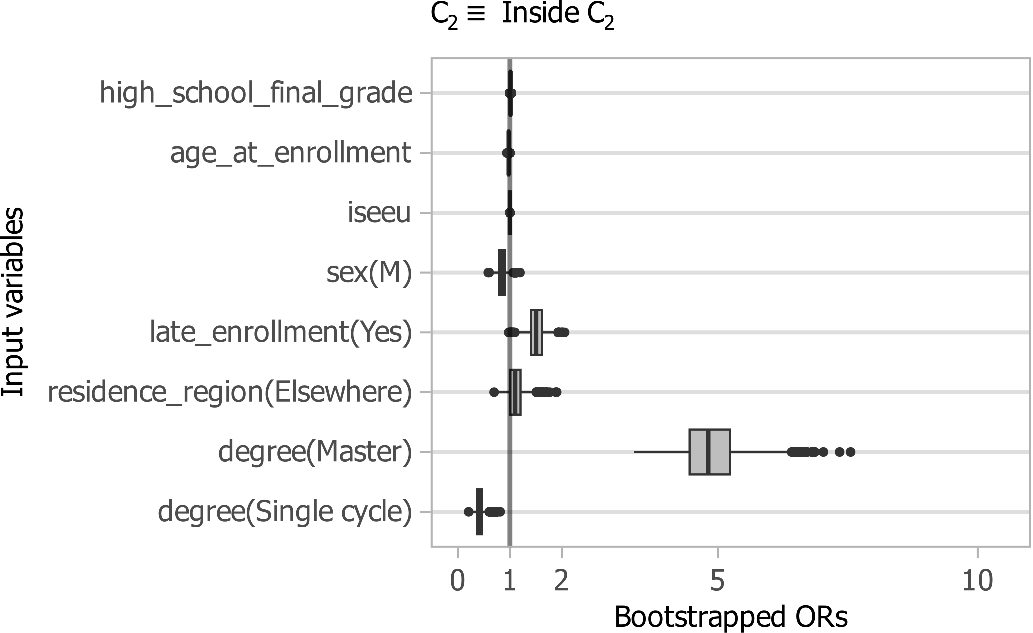}
	\end{center}     
	\caption{\textit{Left panel: Boxplots of the bootstrapped odds ratios on $\mathcal C_1 \equiv$ (outside $\mathcal C_2$) for each input variable, obtained from $B = 3000$ bootstrap samples by estimating the sparse model using the group-LASSO algorithm on each sample.
			Right panel: The only difference in this panel is that the boxplots report the odds ratios on $\mathcal C_2$ (inside $\mathcal C_2$).}}\label{fig:ORSboot}
\end{figure*}

While the aforementioned sub-optimal solution is presented for illustrative purposes only, a comprehensive inferential framework can be achieved through the bootstrap approach detailed in Section \ref{sec:bootstrap}. The boxplots shown in Figure \ref{fig:ORSboot} provide a graphical representation of the distribution of the odds ratios inside and outside $\mathcal C_2$ for each input variable. These distributions were obtained by resampling $B = 3000$ bootstrap samples and estimating the sparse model on each sample, without imposing any constraint on the maximum number of allowable interactions. For each estimated model, the odds ratios were computed directly from the sparse model fitted via the group-LASSO algorithm, using the expressions developed in Subsection \ref{subsec:oddscalculation}. This procedure obviated the need to re-estimate, at each iteration, an additional logistic regression model using the \texttt{glm()} function, which, as illustrated in the previously discussed example, produces solutions that are not perfectly aligned with those of the sparsified model.

The descriptive statistics of the bootstrap estimates of the odds ratios are presented in Table \ref{tab:tab5}. In addition to the mean and standard deviation of the estimates within each cluster, the table also reports the ratios of odds ratios along with their corresponding 95\% BC$_a$ confidence intervals (bias-corrected and accelerated; \citeauthor{diciccio_bootstrap_1996}, \citeyear{diciccio_bootstrap_1996}). As is standard practice, an odds ratio is considered significantly different from 1 if the associated confidence interval does not include 1. The adoption of BC$_a$ bootstrap confidence intervals is motivated by their enhanced robustness in the presence of non-Gaussian characteristics, such as the moderate skewness observed in the distribution of bootstrapped odds ratios shown in Figure \ref{fig:ORSboot}.

\begin{table}[ht]
	\centering\renewcommand*{\arraystretch}{1.0}\caption{\textit{Descriptive summaries of the odds ratios and the ratios of odds ratios for each input variable, estimated from the $B = 3000$ bootstrap replicates of the sparse model. The reported confidence intervals are the 95\% BC$_a$ confidence intervals (bias-corrected and accelerated). An odds ratio is considered significantly different from 1 if the corresponding confidence interval does not include 1 (intervals meeting this criterion are indicated in bold).}}
	\label{tab:tab5}\resizebox{\textwidth}{!}{\begin{tabular}{lccccccccc}
	\hline\hline
	& $\mathcal C_1$(mean) & $\mathcal C_1$(sd) & $\mathcal C_1$(95\% BC$_a$) & $\mathcal C_2$(mean) & $\mathcal C_2$(sd) & $\mathcal C_2$(95\% BC$_a$) & \textsf{ROR}s(mean) & \textsf{ROR}s(sd) & \textsf{ROR}s(95\% BC$_a$)\\
	\hline
	\texttt{high\_school\_final\_grade} & 1.01 & $<$0.01 & \textbf{1.01--1.02} & 1.01 & $<$0.01 & 1.00--1.02 & 1.00 & $<$0.01 & 0.98--1.00\\
	\texttt{age\_at\_enrollment} & 0.96 & 0.01 & \textbf{0.94--0.99} & 0.98 & 0.01 & \textbf{0.96--0.99} & 1.01 & 0.01 & 0.99--1.04\\
	\texttt{iseeu} & 1.00 & 0.00 & 1.00--1.00 & 1.00 & 0.00 & 1.00--1.00 & 1.00 & $<$0.01 & 1.00--1.00\\
	\texttt{\textbf{sex}}(M) & 0.76 & 0.05 & \textbf{0.67--0.87} & 0.85 & 0.08 & 0.72--1.04 & 1.12 & 0.12 & 0.95--1.45\\
	\texttt{late\_enrollment}(Yes) & 1.54 & 0.15 & \textbf{1.29--1.89} & 1.51 & 0.16 & \textbf{1.23--1.84} & 0.99 & 0.06 & 0.54--1.00\\
	\texttt{residence\_region}(Elsewhere) & 1.32 & 0.17 & \textbf{1.06--1.76} & 1.12 & 0.15 & 0.89--1.50 & 0.85 & 0.14 & 0.58--1.11\\
	\texttt{degree}{(Master)} & 4.36 & 0.72 & \textbf{2.85--5.68} & 4.87 & 0.58 & \textbf{3.92--6.25} & 1.14 & 0.20 & 1.00--2.24\\
	\texttt{degree}{(Single cycle)} & 0.43 & 0.06 & \textbf{0.33--0.56} & 0.42 & 0.07 & \textbf{0.29--0.59} & 0.99 & 0.19 & 0.54--1.25\\
	\hline
\end{tabular}}
\end{table}

The results presented in Table \ref{tab:tab5} clearly illustrate that the individual odds ratios and the corresponding ratios of odds ratios serve distinct inferential purposes. While the cluster-specific odds ratios are employed to evaluate the significance of the effect of a given input variable on the probability of dropout, the ratios of odds ratios are designed to assess the significance of differential effects between the two identified clusters. This interpretation extends naturally to cases with $k > 2$ groups, where each cluster is compared against the designated reference cluster.

It is worth noting that some variables are significant in both clusters. This is exemplified by \texttt{high\_school\_final\_grade}, which indicates that the probability of completing university studies increases by approximately 1\% for each additional score in the final grade among students assigned to cluster $\mathcal C_1$. However, this effect is nearly identical in the other cluster, and the corresponding odds ratio is not statistically significant. Conversely, the marked tendency of male students to drop out (relative to female students) is confirmed, with an observed imbalance of approximately 12\%, despite the relative \textsf{ROR} being only marginally significant (becoming significant at the 90\% confidence level). This example illustrates that the statistical significance of the \textsf{ROR} may be considered secondary when the effects within both clusters are strongly significant and reveal a clear imbalance in dropout risk. Moreover, the particularly elevated dropout risk among students enrolled in single-cycle degree programs--compared to those enrolled in Bachelor's programs, likely due to the greater level of commitment required--further underscores that, even the dataset has limited number of variables, the results highlight the need for careful monitoring of male students in cluster $\mathcal C_2$ who enroll in single-cycle programs, as they represent one of the groups most at risk of dropout. The value of the proposed methodological framework becomes even more evident in data-rich contexts, where a greater number of features are available for each student, or in predictive settings involving prospective students.

\section{Discussion and conclusions}\label{sec:discussion}

In this work, we propose a straightforward extension of the standard logistic regression model and apply it to the analysis of student dropout in higher education. This phenomenon is plausibly heterogeneous, and the proposed model exploits a data partition generated by a clustering algorithm, which identifies subpopulations that are homogeneous not only with respect to the feature variables but also in terms of the outcome variable of interest (dropout or graduation). The resulting cluster labels are subsequently incorporated as covariates in a logistic regression model that includes interaction terms between these labels and each main effect included in the model. We develop the theoretical framework necessary to correctly interpret the parameters of this extended model and to perform valid inference, even after model simplification via a specific variant of the LASSO algorithm that enforces the strong hierarchy property.

The proposed methodology follows a two-stage data analysis pipeline. The first stage is unsupervised and aims to capture latent heterogeneity by identifying subpopulations of students. This is followed by a supervised stage, in which we estimate our specialized logistic regression model, which can, of course, also be employed for predictive purposes on future instances. One may argue that this separation into two distinct stages is artificial or unnecessary. For example, a class of mixture models known as Mixture of Experts (MoE) provides a unified framework that can address both stages simultaneously \citep{yuksel_twenty_2012,nguyen_practical_2018}. Instances in the training set can be partitioned into clusters, and by using the gating function--that is, mixture weights expressed as functions of the covariates in a multinomial logistic regression model--it becomes possible to allocate future students to the appropriate subgroup. Simultaneously, the gating function determines a distinct logistic regression model for each subpopulation.

Although this approach is conceptually coherent, the estimation of a MoE model is particularly challenging from a computational standpoint, as is the case for mixture models in general. Moreover, MoEs are susceptible to overfitting and are notoriously difficult to fine-tune, which hinders their practical applicability--especially when the number of mixture components is unknown and must be selected from the training data \citep{celeux2018model}. In contrast, the procedure we propose introduces no free parameters, apart from selecting the number of groups in the clustering algorithm--a task that is typically much simpler than determining the number of components in a finite mixture model. In any case, the partition may be determined by an expert data scientist, and the final output consists of easily interpretable odds ratios. This interpretability is preserved even in high-dimensional settings involving a large number of input variables, and the resulting estimates can be directly leveraged to inform and guide the development of targeted policies aimed at addressing student dropout.

As previously noted, a limitation of the current version of the theory arises from the inherent nature of the data employed. In the logistic regression model utilized, the variable representing the clustering label is a nonlinear function--through the clustering algorithm--of all other variables, including the response variable $Y$. Consequently, when applying resampling procedures or cross-validation, the functional/nonlinear relationship between the variables and the partition labels, as captured by the clustering algorithm in the original training data, is preserved. This may result in an underestimation of the generalization error, a phenomenon that remains unquantified at present. An alternative approach would involve determining a new partition each time the data is resampled or a fold is randomly extracted. While this method is expected to provide more conservative estimates of the generalization error, it introduces a significant increase in the computational cost. In contrast, the simplified version of the data analysis pipeline we propose is computationally more efficient. However, we are currently unable to quantify its impact on the expected test error.

As previously highlighted in Subsection \ref{sec:standard}, our model is expected to demonstrate superior predictive performance for prospective students when compared to a standard logistic regression model. A thorough comparative analysis of predictive accuracy, benchmarked against various machine learning algorithms commonly employed in the literature for forecasting university dropout \citep{vaarma_predicting_2024}, represents a promising avenue for future research to further validate our proposed approach.

In conclusion, we contend that the proposed approach represents a valuable addition to the existing methods for analyzing university dropout data. It provides easily interpretable effect sizes, based on odds ratios, which inherently account for the latent heterogeneity within the university student population, particularly regarding the propensity for premature dropout.


\section*{Declaration of Conflicting Interests}
The authors declare that they have no actual or potential conflicts of interest regarding the research, authorship, or publication of this article.

\section*{ORCID IDs}

Andrea Nigri \orcidlink{https://orcid.org/0000-0002-2707-3678}{ https://orcid.org/0000-0002-2707-3678} \\
Massimo Bilancia \orcidlink{https://orcid.org/0000-0002-5330-2403}{ https://orcid.org/0000-0002-5330-2403} \\
Barbara Cafarelli \orcidlink{https://orcid.org/0000-0002-7385-4213}{ https://orcid.org/0000-0002-7385-4213} \\
Samuele Magro \orcidlink{https://orcid.org/0009-0000-5846-4674}{ https://orcid.org/0009-0000-5846-4674}

\section*{Acknowldgments}

A.~Nigri, and B.~Cafarelli acknowledge financial support under the National Recovery and Resilience Plan (NRRP), Mission 4, Component 2, Investment 1.1, Call for tender No.~1409 published on 14.9.2022 by the Italian Ministry of University and Research (MUR),funded by the European Union -- Next Generation EU -- Project Title: \textit{Evaluation strategies to contrast university drop-out through empowerment and to design educational and career guidance practices} -- CUPD53D23020530001 -- Grant Assignment Decree No.~P2022XYN9A, by the Italian Ministry of University and Research (MUR). \\
\noindent M.~Bilancia acknowledges financial support under a research grant from the University of Study of Foggia -- Project Title: \textit{Application of Machine Learning tools for the prediction of university dropout risk} (Official University Register, protocol no.~05436 – VII/1 of 03/02/2025, record no.~182/2025 – AUA 239/2025).

\end{document}